\documentclass[twocolumn]{aastex631}
\pdfoutput=1

\usepackage{natbib}
\usepackage{graphicx}
\usepackage{hyperref}
\usepackage{graphicx}
\usepackage{float}

\usepackage{siunitx}
\usepackage{amsmath}
\usepackage{appendix}
\usepackage{placeins}
\usepackage{bm}

\newcommand{\uveci}{{\bm{\hat{\textnormal{\bfseries\i}}}}}
\newcommand{\uvecj}{{\bm{\hat{\textnormal{\bfseries\j}}}}}

\DeclareSIUnit \parsec {pc}
\DeclareSIUnit \year {yr}

\shorttitle{Gaseous Distribution Morphology}
\shortauthors{Waters et al.}

\accepted{in ApJ}

\begin{document}

\title{Gas Morphology of Milky Way-like Galaxies in the TNG50 Simulation: Signals of Twisting and Stretching}

\correspondingauthor{Thomas K.\ Waters}
\email{waterstk@umich.edu}

\author[0000-0002-5231-7240]{Thomas K.\ Waters}
\affiliation{University of Michigan $\vert$ Department of Astronomy, 1085 S University, Ann Arbor, MI 48109, USA}
\affiliation{University of Washington $\vert$ Departments of Physics \& Astronomy, 1400 NE Campus Parkway, Seattle, WA 98195, USA}

\author[0000-0003-3342-5286]{Colton \ Peterson}
\affiliation{University of Washington $\vert$ Departments of Physics \& Astronomy, 1400 NE Campus Parkway, Seattle, WA 98195, USA}

\author[0000-0002-2791-5011]{Razieh \ Emami}
\affiliation{Center for Astrophysics $\vert$ Harvard \& Smithsonian, 60 Garden Street, Cambridge, MA 02138, USA}

\author[0000-0002-6196-823X]{Xuejian \ Shen}
\affiliation{TAPIR, California Institute of Technology, Pasadena, CA 91125, USA}

\author[0000-0001-6950-1629]{Lars \ Hernquist}
\affiliation{Center for Astrophysics $\vert$ Harvard \& Smithsonian, 60 Garden Street,  Cambridge, MA 02138, USA}

\author[0000-0003-4284-4167]{Randall \ Smith}
\affiliation{Center for Astrophysics $\vert$ Harvard \& Smithsonian, 60 Garden Street, Cambridge, MA 02138, USA}

\author[0000-0001-8593-7692]{Mark \ Vogelsberger}
\affiliation{Department of Physics, Kavli Institute for Astrophysics and Space Research, Massachusetts Institute of Technology, Cambridge, MA 02139, USA}

\author[0000-0002-7892-3636]{Charles \ Alcock}
\affiliation{Center for Astrophysics $\vert$ Harvard \& Smithsonian, 60 Garden Street, Cambridge, MA 02138, USA}

\author[0000-0002-5445-5401]{Grant \ Tremblay}
\affiliation{Center for Astrophysics $\vert$ Harvard \& Smithsonian, 60 Garden Street, Cambridge, MA 02138, USA}

\author[0000-0003-4475-9345]{Matthew \ Liska}
\affiliation{Center for Astrophysics $\vert$ Harvard \& Smithsonian, 60 Garden Street, Cambridge, MA 02138, USA}

\author[0000-0002-1975-4449]{John C. \ Forbes}
\affiliation{Center for Computational Astrophysics, Flatiron Institute, New York, USA}

\author[0000-0002-3430-3232]{Jorge \ Moreno}
\affiliation{Department of Physics and Astronomy, Pomona College, Claremont, CA 91711, USA}

\begin{abstract}

\noindent We present an in-depth analysis of gas morphologies for a sample of 25 Milky Way-like galaxies from the IllustrisTNG TNG50 simulation. We constrain the morphology of cold, warm, hot gas, and gas particles as a whole using a Local Shell Iterative Method (LSIM) and explore its observational implications by computing the hard-to-soft X-ray ratio,  which ranges between $10^{-3}$-$10^{-2}$ in the inner $\sim 50 \rm kpc$ of the distribution and $10^{-5}$-$10^{-4}$ at the outer portion of the hot gas distribution. We group galaxies into three main categories: simple, stretched, and twisted. These categories are based on the radial reorientation of the principal axes of the reduced inertia tensor. We find that a vast majority ($77\%$) of the galaxies in our sample exhibit twisting patterns in their radial profiles. Additionally, we present detailed comparisons between 1) the gaseous distributions belonging to individual temperature regimes, 2) the cold gas distributions and stellar distributions, and 3) the gaseous distributions and dark matter (DM) halos. We find a strong correlation between the morphological properties of the cold gas and stellar distributions. Furthermore, we find a correlation between gaseous distributions with DM halo that increases with gas temperature, implying that we may use the warm-hot gaseous morphology as a tracer to probe the DM morphology. Finally, we show gaseous distributions exhibit significantly more prolate morphologies than the stellar distributions and DM halos, which we hypothesize is due to stellar and AGN feedback. 

\end{abstract}


\section{Introduction} \label{sec:intro}

    Standard galactic morphology classification, with methods such as the Hubble sequence \citep{Hubble1, Hubble2, Hubble3, Hubble4}, the Hubble Comb \citep{hubblecomb}, and the de Vaucouleurs system \citep{DeVal}, differentiates galaxies based on their physical appearance in optical bands merely by tracing the main-sequence stellar distribution. Galaxies are primarily separated into elliptical, spiral, barred spiral, and irregular morphologies. In addition to the above morphologies, the galaxies are also surrounded by roughly spherical halos consisting of older Population II stars \citep{PopII}, diffuse gas with a wide range of temperatures, and dark matter, each displaying their own unique morphological properties \citep{DMHaloMorph, StarMorph}. However, unlike galactic stellar morphology, dark matter and gas shapes must be gleaned from more inferential methods and cannot be directly observed from optical data alone \citep{CGM}.
    
    The properties of the galactic gas in the Milky Way (MW) have been probed using several methods. \citet{structure_MW_halo} examine the structure of the MW's hot gas halo using the {\sl XMM-Newton} Reflection Grating Spectrometer archival data. They measure the absorption line strengths along the line of sight in several AGN and extra-galactic sources. \citet{MW_halo_emission} later refined their initial constraints on the MW's hot gas mass by examining multiply ionized oxygen emission lines with a much larger sample size. Using this technique, they report hot gas masses of $M(<50\, {\rm kpc}) = 3.8^{+0.3}_{-0.3} \times 10^9 M_\odot$ and $M(<250 \, {\rm kpc}) = 4.3^{+0.9}_{-0.8} \times 10^{10} M_\odot$. To constrain the hot gas halo density distribution, \citet{MW_density_pulsars} examined the dispersion measure of pulsars with independently known distances. \citet{MW_gas_spacial_distrib} used X-ray data from {\sl Suzaku} to constrain the spatial distribution of the hot gas morphology for the MW, which consists of a dense disky component and a diffuse spherical component. \citet{2021ApJ...912....9W} studied the morphology of the Circumgalactic Medium (CGM) using COS + Gemini maps in a sample of 1689 galaxies. From this, they estimated the radial extent of the CGM using the HI covering fraction. \cite{2022ApJ...927..147T} studied the CGM of 126 galaxies with O VI from the CGM2 survey and made some test sets on different CGM models. The properties of stellar distributions and dark matter halos have also been probed observationally. For example, \citet{bertola} studied ellipticity and isophote twisting in elliptical galaxies and \citet{allen} used observations from the \textit{Chandra} observatory to constrain mass–temperature, and temperature–luminosity relations, which confirmed a $40\%$ offset between observed and predicted mass distributions.

    On the theoretical front, there have been some studies using different hydrodynamical simulations to constrain the structures of hot gaseous halos, stellar distributions, and dark matter halos. \citet{CGM_morphology} examined the CGM of MW-like galaxies in the Illustris and the IllustrisTNG simulations to probe feedback physics. \citet{2019MNRAS.488.1248H, 2020MNRAS.494.3581H} probed the origin of the CGM from the FIRE simulation. \citet{2021MNRAS.500.1038L} also used FIRE simulations and studied the structure and composition of the gas in CGM in low redshift dwarf galaxies. \citet{IGM} used six hydrodynamical simulations to study the intergalactic medium (IGM) in the temperature regime of $10^5 < T < 10^7 K$. To probe the structure of dark matter halos, \citet{bonamigo} employed the Millennium XXL (a cosmological dark matter-only simulation) and SBARBINE simulations to determine the major and intermediate to major axis ratio distributions of a wide range of dark matter halo masses. \citet{pulsoni} explored the shape of stellar halos of 1114 early-type galaxies (ETGs) in both the IllustrisTNG TNG100 and TNG50 simulations, finding that their sample widely represents observed kinematic properties and intrinsic stellar halo shapes of ETGs.
    
    Here, we examine the morphologies of galactic gas in different temperature regimes in a sample made of 25 Milky Way-like galaxies in the TNG50 simulation \citep{TNGDataRelease, TNGFirstResultsFB, TNGFirstResultsEV}, which displays the highest resolution in the series of IllustrisTNG simulations. In particular, we constrain the gas morphologies for cold, warm, hot, and all gas (as defined in Table~\ref{tab: tempclass}) in each of these 25 galaxies. We infer the radial profile of the galactic gas shape using a Local Shell Iterative Method (LSIM) \citep{StarMorph, DMHaloMorph}. In each temperature regime, we classify galaxies in the following categories: simple, twisted, stretched, and twisted-stretched. Finally, we compare galaxy morphologies between individual temperature regimes as well as the dark matter and stellar distributions.

    The paper is organized as follows: Section~\ref{sec: method} outlines the MW-like galaxy sample selection, temperature calculation and cuts, and observational implications. Section~\ref{sec: shape_analysis} presents our analysis of the shape of the gas particles from the TNG50 simulation. Section~\ref{sec: shape_class} presents our classifications and definitions of the various types of gaseous distributions. Section~\ref{sec: comparison} compares gaseous distribution morphologies in various temperature regimes with the dark matter halos and stellar distributions. Section~\ref{sec: observations} puts this work in the context of current and future observations and similar works with hydrodynamic simulations. Finally, Section~\ref{sec: conclusions} summarizes the work, outlines our results, and explores implications and future work. In addition, we comment on the arbitrary nature of eigenvector orientation in Appendix~\ref{appendix:eigenvectors} and provide some supplementary figures in Appendix~\ref{appendix:viz} to make our classification scheme more intuitive.
    
    For this work, we adopt the standard TNG50 values for the following cosmological parameters \citep{TNG_parameters}: $\Omega_M = \Omega_{dm} + \Omega_b = 0.3089$, $\Omega_b = 0.0486$, $\Omega_{\Lambda}= 0.6911$, $H_0 = 100h$ \si{\km\per\second\per\mega\parsec}, $h = 0.6774$, $\sigma_8 = 0.8159$, and $n_s = 0.9667$.  
  
\begin{figure*}
\centering
\plotone{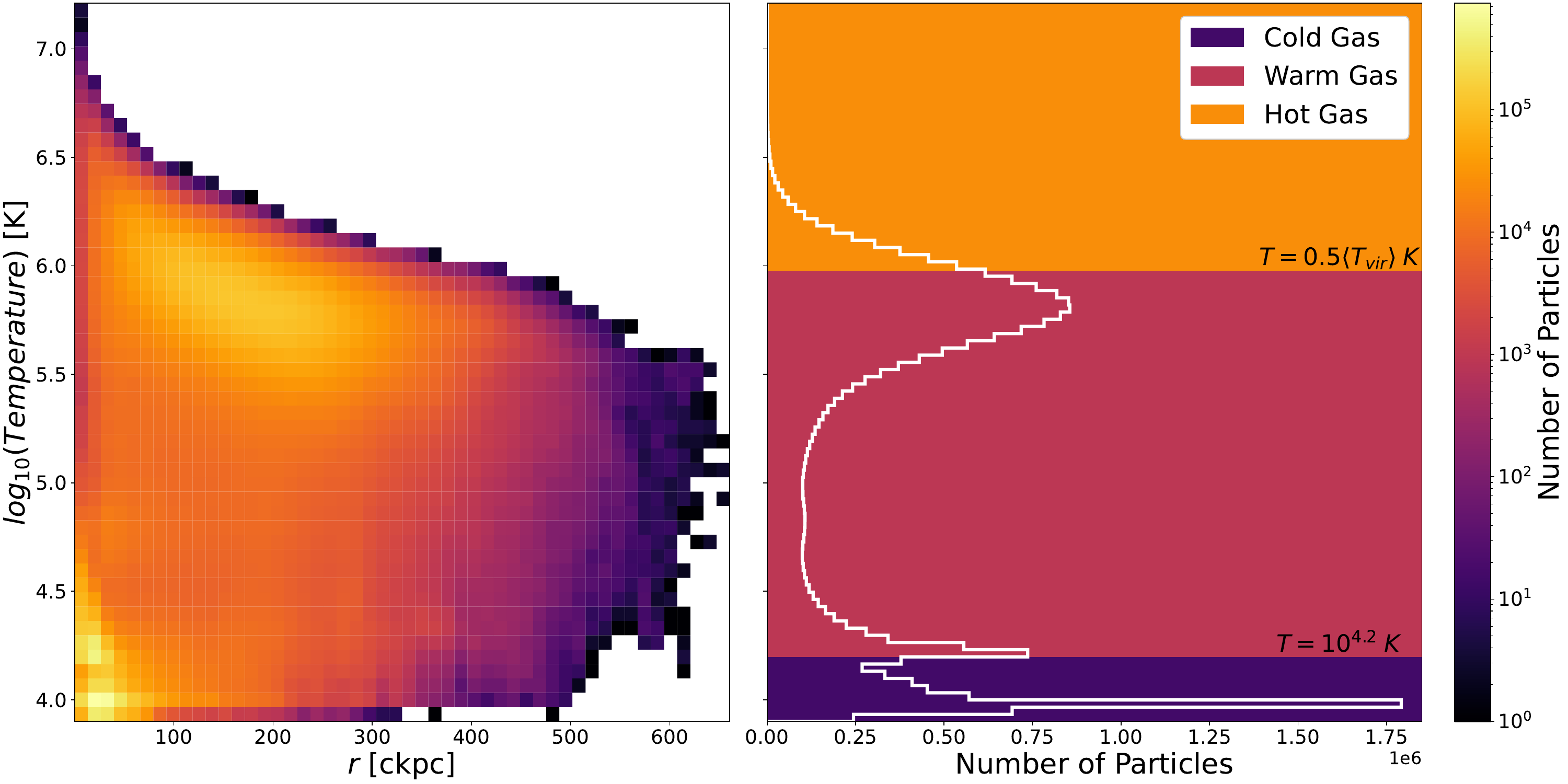}
\caption{Temperature distributions of the cumulative gas particles from all 25 galaxies in our sample. The left panel displays a two-dimensional histogram showing the logarithm of the temperature of the gas particles as a function of the galactic radii. The right panel displays a one-dimensional trimodal distribution of the logarithm of the temperature of the gas particles. Overlaid on these distributions are the temperature demarcations for cold ($T<10^{4.2}K$), warm ($10^{4.2}K<T<$ $0.5T_{\rm vir}$), and hot ($T>$ $0.5T_{\rm vir}$) gas in purple, red, and orange respectively. The color bar indicates the number of particles at a given temperature and radius for the left panel and is scaled logarithmically.} 
\label{fig: temp_1d_2d}
\end{figure*} 
 
\begin{figure*}[ht!]
\centering
\includegraphics[width = 0.8\paperwidth]{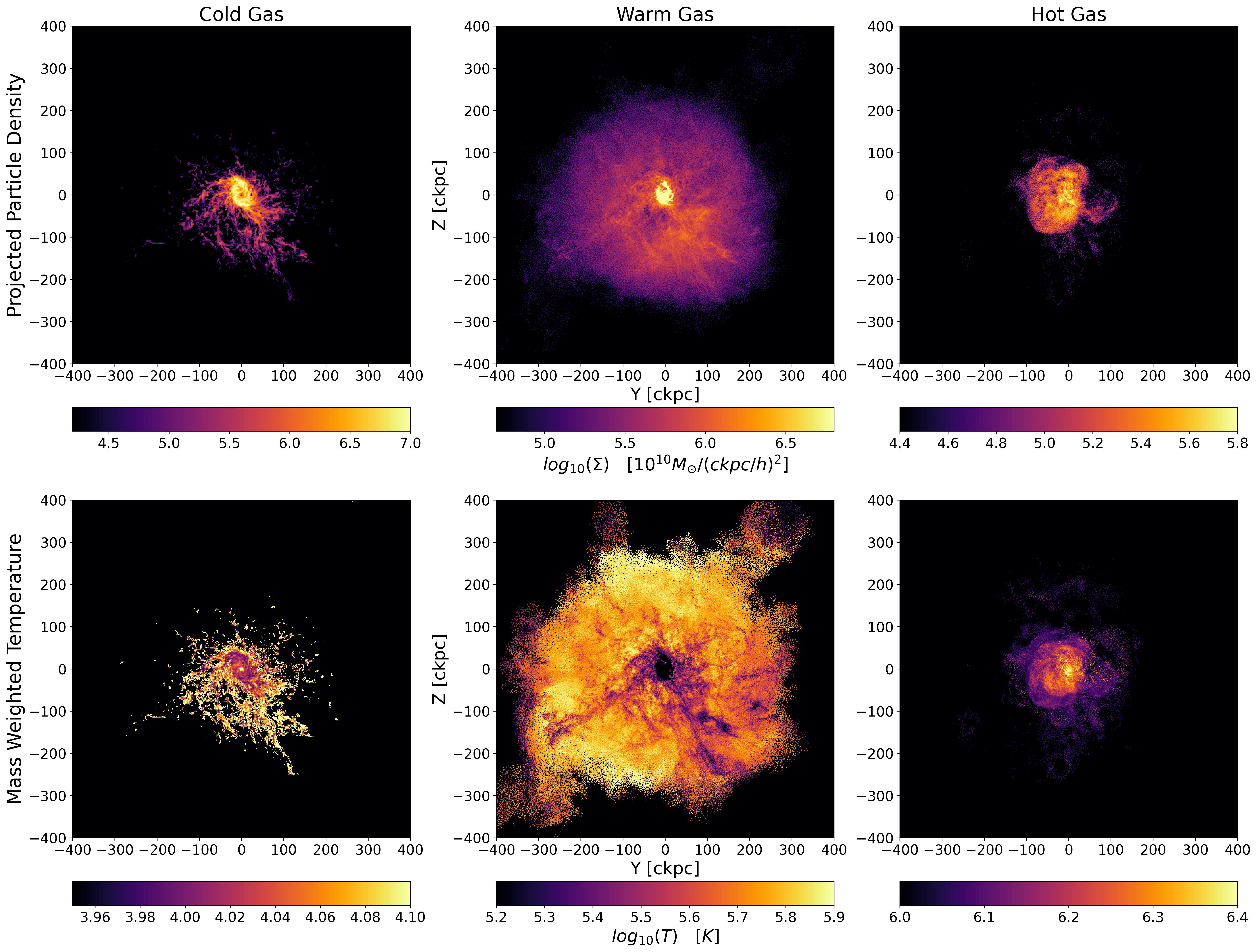}
\caption{Projected surface density ($\Sigma$, top panels) and mass weighted temperature ($T$, bottom panels) maps of the YZ projection for the MW-like galaxy with ID number 479938 in our sample of 25 galaxies illustrated with its cold gas ($T < 10^{4.2}K$) in the left panels, warm gas ($10^{4.2}K < T <$ $0.5T_{\rm vir}$) in the middle panels, and the hot gas ($T > $ $0.5T_{\rm vir}$) in the right panels. The projected particle density and mass-weighted temperature are given on logarithmic scales.
\label{fig: density_maps}}
\end{figure*}

\section{Methodology} \label{sec: method}
    
    Below, we briefly introduce the TNG50 simulation \citep{TNGFirstResultsEV, TNGFirstResultsFB} and illustrate our methodology for making a sample of MW-like galaxies, following the approach chosen in \cite{StarMorph, DMHaloMorph, 2022arXiv220207162E}. We also present temperature calculations for gas particles and separate them into three temperature regimes: cold, warm, and hot gas. Finally, we explore observational implications by presenting a brief exploration of the X-ray luminosity.

\subsection{TNG50 Simulation} \label{sec: TNG50}

    The TNG50 simulation is a cosmological, magnetohydrodynamical simulation used to simulate galaxy formation and evolution. This simulation is the third in the series of IllustrisTNG projects and evolves $2160^3$ dark matter particles and gas cells in a periodic box. It also evolves stars, magnetic fields, and supermassive black holes using a periodic boundary condition. The volume of the TNG50 is $51.7^3$ cMpc$^3$ with a softening length of $0.39$ c\si{\kilo\parsec\per\hour} for $z \geq 1$ and $0.195$ proper \si{\kilo\parsec\per\hour} for $z < 1$. Consequently, TNG50 exhibits a high numerical resolution comparable to zoom-in simulations \citep{TNGFirstResultsEV}.  
        
    The initial conditions were chosen at $z=127$ using the Zeldovich approximation and the simulation is selected from a set of 60 random iterations of the initial density field. The initial conditions are evolved using the AREPO code \citep{AREPO}. Finally, Poisson's equations are solved with a tree-particle-mesh algorithm. Furthermore, the TNG50 models for unresolved astrophysical processes are identical to the other IllustrisTNG simulations and are described in \citet{Weinberger2017} and \citet{unresolved}.

\subsection{MW-like galaxy Sample Selection} \label{sec: sample}

    To choose our sample of MW-like galaxies, we follow the procedures outlined in \citet{StarMorph,DMHaloMorph}. In the first step, we select galaxies with dark matter halo masses in the range $(1-1.6)\times10^{12} M_\odot$. This subset contains a total of 71 galaxies. We then narrow this subset further by choosing only the galaxies that have more than 40\% of their stellar populations being rotationally supported. This selection is made by implementing the orbital circularity parameter shown below:    
    
\begin{equation} \label{eq: circularity}
\varepsilon_i \equiv \frac{j_{z,i}}{j_c(E_i)}, \quad j_c(E_i) = r_cv_c = \sqrt{GM(<r_c)r_c}. \quad
\end{equation}
\newline
    where $j_{z,i}$ is the $z$-component of the angular momentum of the $i$-th particle, $j_c(E_i)$ corresponds to the circular angular momentum with radius $r_c$ and energy $E_i$, $v_c$ is the orbital velocity, $G$ is the gravitational constant, and $M(<r_c)$ refers to the mass within $r_c$. To limit our galaxy sample to those which have $>40\%$ of stellar particles, we select galaxies with $\varepsilon_i \geq 0.7$. This ensures us that a substantial portion of stars in the galaxy resides in the disk, resulting in a final sample size of $25$ MW-like galaxies.

\subsection{Gas Temperature Cuts} \label{sec: temp}

    Having presented the sample of MW-like galaxies in the TNG50 simulation, here we compute the gas temperature profile for each galaxy in our sample. As we are interested in examining the morphology of the gaseous distributions in different temperature regimes (cold, warm, and hot gas), we compute the temperature of the gas particles using the following relation:
    
\begin{equation} \label{eq: temp}
T = (\gamma-1)\frac{u}{k_B}\mu,
\end{equation}
    where $T$ is the temperature in Kelvin, $\gamma$ ($= 5/3$) is the adiabatic index, $u$ is the internal energy per unit mass, $k_B$ is the Boltzmann constant in CGS units, and $\mu$ is the mean molecular weight. Here, we use the following definition of the mean molecular weight:
\begin{equation} \label{eq: meanmolecularweight}
\mu = \frac{4}{1+3X_H+4X_H\chi_e} m_p,
\end{equation}
    with $X_H$ ($=0.76$) referring to the hydrogen mass fraction, $\chi_e$ ($=n_e/n_H$) is the electron abundance, and $m_p$ is the mass of the proton.

    To avoid arbitrary temperature demarcations, we define the boundaries between gas temperature regimes with physically motivated quantities. The boundary between the warm and hot gas is defined with respect to the virial temperature (many collaborations define the temperature of the hot gas to be equal to, or some fraction of, the virial temperature. For example, see: \citet{warm_tvir} who define hot gas to have temperatures above $0.5T_{\rm vir}$ and \citet{warmhot_MW} who define their hot gas to be at the virial temperature ($\sim 10^{6} \rm K$). Hereafter, we refer to the virial temperature as $T_{\rm vir}$, which is adapted from \citet{Tvir} and is defined below:
\begin{equation} \label{eq: virialtemp}
    T_{\rm vir}  = \frac{2}{3}\frac{GM_{\rm 200}}{R_{\rm 200}}\frac{\mu}{k_B}.
\end{equation}
    where $M_{\rm 200}$ is the total mass enclosed in a sphere with a mean density being a factor of 200 above the critical density of the universe at $z=0$, $R_{\rm 200}$ is the radius of a sphere with a mean density being a factor of 200 above the critical density of the universe, $\mu$ is the mean molecular weight and $m_H$ is the mass of the hydrogen atom. We compute the virial temperature for each galaxy in our sample separately, so the resultant hot cut is unique to each galaxy.
    
    To define cold gas, we select particles with a 
    temperature less than the H\,\footnotesize II \normalsize recombination temperature ($T \approx 10^{4}K$) \citep{HII_recombination}. Here, we assume that this value is approximately $10^{4.2}K$. This value is selected from Figure~\ref{fig: temp_1d_2d} where the distribution shows a sharp drop off after a local maximum at $\sim 10^{4.22}K$. It should be noted that the cold gas cut in Figure~\ref{fig: temp_1d_2d} lies slightly above the local minimum in the right panel. This figure is composed of all galaxies in our sample, but the temperature distributions of each galaxy are unique. We find that the average local minimum across all galaxies in our sample lies at $10^{4.193}K$. Therefore, we round our cold gas selection criterion to $T<10^{4.2}K$, which does not result in a significant increase in cold gas particles.
    
    For warm gas, we choose particles with temperatures greater than the H\,\footnotesize II \normalsize recombination temperature and less than a factor of 0.5 of $T_{\rm vir}$, analogously to \citep{warm_tvir}. Despite this warm gas cut not occurring at a local minimum, which would be the ideal method to separate gas regimes, our definition of warm gas is in broad agreement with the literature. However, cuts between cold, warm, and hot gas often differ widely in the literature (both in theoretical and observational works). Examples of these criteria include, but are not limited to, the following: \citet{warm_tvir} (theoretical) defines warm gas as gas with temperatures $10^{4.5} < T < 0.5T_{\rm vir}$, \citet{warmhot} (observational) defines a warm-hot phase ranging between $10^6K$ and $10^7 \rm K$, \citet{warmhot_MW} (observational) state that a warm-hot phase of $10^{6.3} \rm K$ and a hot phase to be $10^{6.88} \rm K$ is required to reproduce observations, and \citet{cen_ostriker99}, a widely cited work, define warm gas to be $10^5 < T < 10^7 \rm K$. 
    
    With the lack of a robust physical differentiation between warm and hot gas phases, our cut at $0.5T_{\rm vir}$ could result in similarities in the classifications and analyses of the warm and hot gas. In addition, direct comparisons to observational works may be difficult as there is no standard cut between warm and hot gas in the literature. However, our cut at $0.5T_{\rm vir}$ was chosen to align as closely as possible with the literature and to maximize the convergence of the Local Shell Iterative Method (LSIM) for both warm and hot gas by ensuring that there is a sufficient number of particles in each regime. Specifically, there must be at least $10^3$ particles per shell for LSIM to converge. We also performed LSIM on gas chosen with warm-hot cuts with factors of $0.2$, $0.3$, $0.8$, and $1.0$ of the virial temperature. We found that $0.5T_{\rm vir}$ was the ideal temperature cut as it resulted in the convergence of LSIM in all halos in both the warm and hot gas regimes. However, we find that, when compared to a hot cut of $0.3T_{\rm vir}$, the convergence of LSIM in each halo is limited to a smaller spatial extent ($\sim 10$ to $>200 \rm \, ckpc$ on average for $0.3T_{\rm vir}$ and $\sim 10$ to $150 \, \rm ckpc$ on average for $0.5T_{\rm vir}$). This may result in a less robust hot gas-to-dark matter comparison. Since both the virial temperature and the overall particle temperature distributions are unique to each galaxy, making a cut at a factor of 0.5 of the virial temperature also ensures a more consistent warm/hot gas particle count than a hard temperature cut. The convergence criteria are outlined in more detail in Section~\ref{sec: LSIM}. Finally, for hot gas, we adopt particles with temperatures greater than $0.5 T_{\rm vir}$. 

    Figure~\ref{fig: temp_1d_2d} shows the 2D temperature histogram as a function of galactic radius (left) and the 1D distribution of the gas temperature (right) for all 25 Milky Way-like galaxies in our sample, respectively. The right panel displays a trimodal temperature distribution with our temperature demarcations overlaid on the plot where cold, warm, and hot gas are shown with purple, red, and orange regions, respectively. 
    
    The first peak from the bottom on the right panel refers to the cold, star-forming, gas particles located mostly in the galactic disk. This is followed by the second peak describing H\,\footnotesize II \normalsize regions, and finally, the third peak corresponds to the warm--hot gas. As can be seen in Figure~\ref{fig: temp_1d_2d}, a hot cut of $0.5 T_{\rm vir}$ provides a sufficient number of particles in both the warm and hot temperature regimes. Table~\ref{tab: tempclass} summarizes the above temperature cuts across different regimes. 
    
    Figure~\ref{fig: density_maps} presents three projections of the cold, warm, and hot gas particles for one galaxy in our sample (with ID number 479938). In the top row, we show the surface density. In the bottom row, we present mass-weighted temperature. The difference in the spatial distribution of gas particles is evident by examining Figure~\ref{fig: density_maps} where the definition of the spiral structure in the galaxy becomes less apparent as the temperature is increased beyond the cold gas threshold (i.e. $T>10^{4.2}K$).  

\begin{table}[ht]
\centering
\caption{Gaseous classification in our MW-like galaxy sample.} \label{tab: tempclass}
\begin{tabular}{ll}
\hline
\hline
\multicolumn{1}{l}{Gas Classification} & \multicolumn{1}{l}{Temperature Criterion} \\
\hline \hline
Cold               & $T < 10^{4.2}K$          \\ \hline
Warm               & $10^{4.2}K < T < $$\,0.5T_{\rm vir}$ \\ \hline
Hot                & $T > $$\,0.5T_{\rm vir}$  \\
\hline
\end{tabular}
\end{table}  

\begin{figure}
\centering
\includegraphics[width=0.49\textwidth]{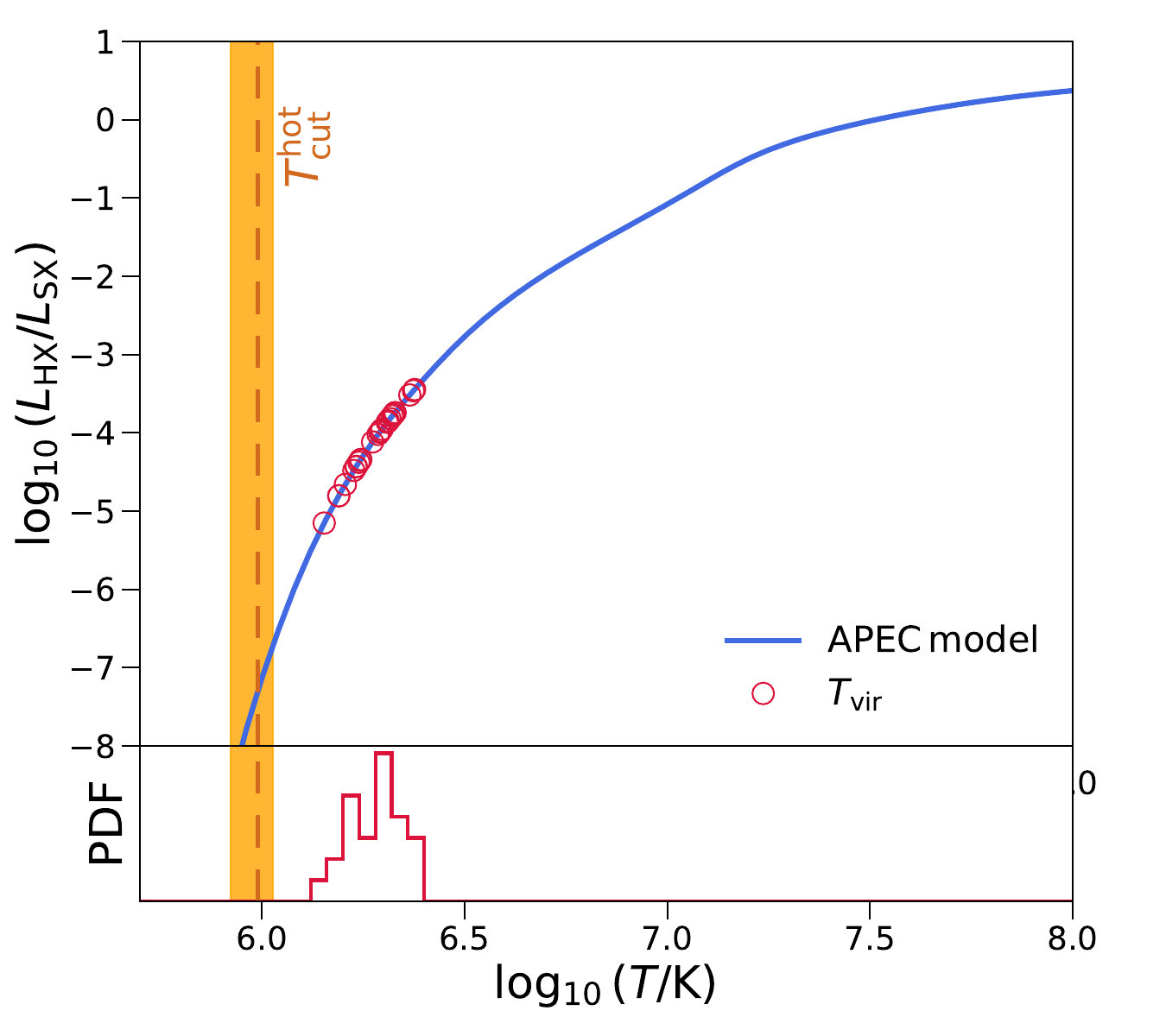}
\caption{ Hard-to-soft X-ray luminosity ratio as a function of gas temperature. The X-ray fluxes are calculated using the APEC model (see the main text) assuming that the hot gas is in collisional-ionization equilibrium. The virial temperatures of the $25$ MW-like galaxies in the sample are shown as red points, and the distribution function of virial temperatures is shown in the bottom sub-panel. The hot gas temperature criteria $T^{\rm hot}_{\rm cut} \equiv$ $0.5\,T_{\rm vir}$ are shown as the vertical orange dashed line (median) with the shaded region showing $1\sigma$ dispersion. }
\label{fig: hx-sx-ratio}
\end{figure}

\subsection{X-Ray luminosity} \label{sec: Xray}
    To explore observational implications, we compute the hard-to-soft X-ray luminosity ratios for the hot gas identified in simulations. The X-ray spectral templates are generated using the Astrophysical Plasma Emission Code \citep[APEC][]{Smith2001} model implemented in the {\sc PyAtomDB} code\footnote{\href{https://atomdb.readthedocs.io/en/master/index.html}{{\sc PyAtomDB} documentation.}}, which utilizes the atomic data from AtomDB v3.0.9 \citep[last described in][]{Foster2012}. The model gives the emission spectrum of a collisionally ionized diffuse gas in equilibrium with a given temperature and metal abundance pattern. The temperature of gas cells in simulations is calculated following Eq.~\ref{eq: temp}. For simplicity, the abundance pattern is set to solar values following \citet{AG1989} and the intracluster medium metallicity is set to $0.25$~\citep[e.g.,][]{McDonald2016,Mantz2017}. For the hot intracluster gas considered here, the emission is dominated by thermal Bremsstrahlung and the broad X-ray band flux is insensitive to the details of the abundance pattern. Then, we account for Galactic absorption with the photoelectric absorption cross section given by \citet{Morrison1983}, assuming a fixed galactic hydrogen column density of $N_{\rm H} = 2\times 10^{20}\,{\rm cm}^{-2}$.

\begin{figure}[ht!]
\centering
\includegraphics[width = 0.38\paperwidth]{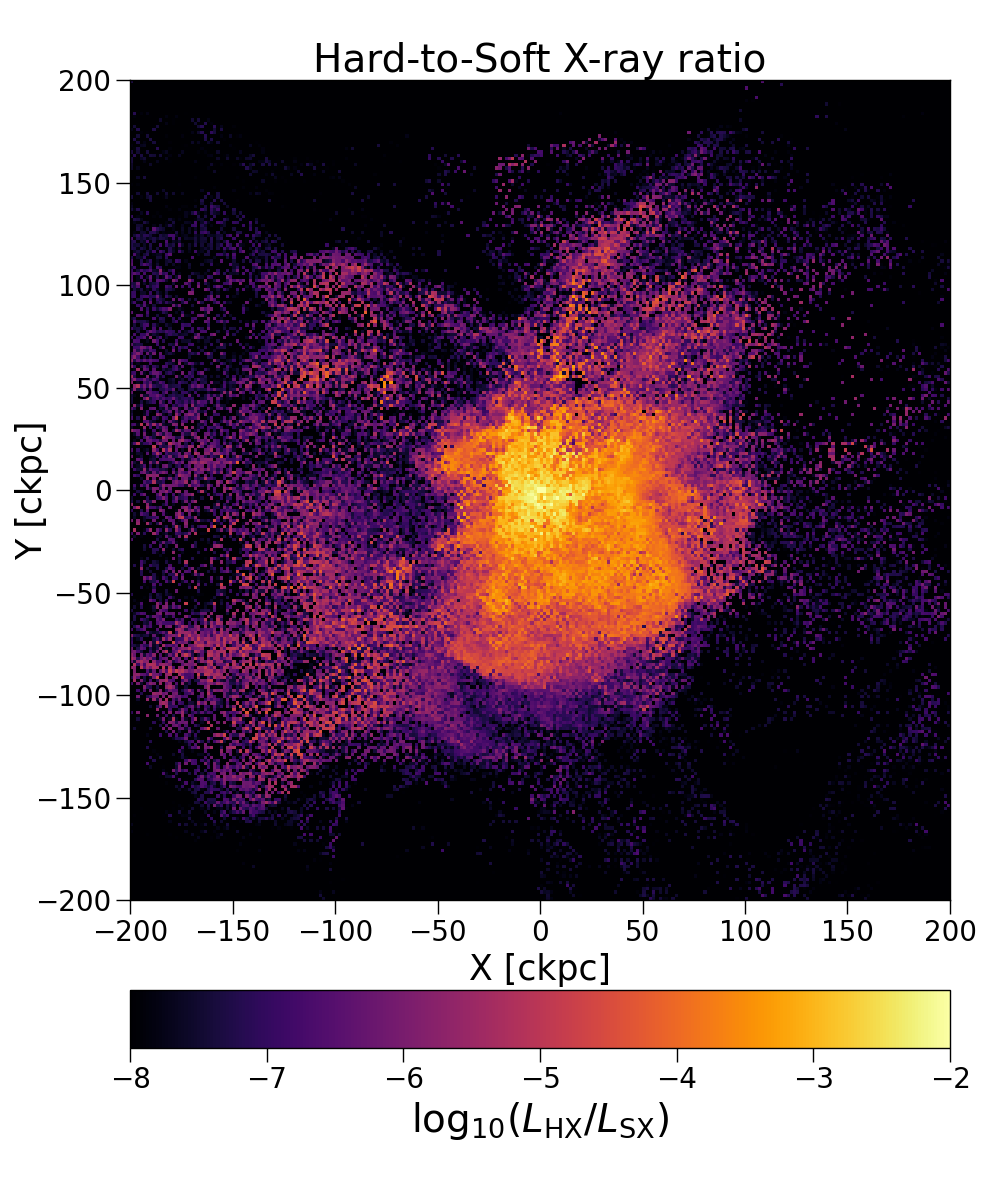}
\caption{Map of the ratio of the projected surface brightness in hard X-ray versus soft X-ray. For illustrative purposes, the map is created for the MW-like galaxy with ID number 479938 in our sample of 25 galaxies.
\label{fig: hx-sx-ratio_maps}}
\end{figure}

    In Figure~\ref{fig: hx-sx-ratio}, we show the hard-to-soft X-ray luminosity ratio as a function of gas temperature. The luminosity ratio decreases monotonically as temperature drops due to the movement of the cut-off energy of the thermal Bremsstrahlung spectrum. The virial temperatures of the chosen MW-like galaxies are typically in the range $10^{6}-10^{6.5}\,{\rm K}$, which implies a typical hard-to-soft X-ray luminosity ratio $\sim 10^{-4}$. Meanwhile, the temperature criterion for hot gas, $T^{\rm hot}_{\rm cut} \equiv$ $0.5\,T_{\rm vir}$, is appropriately chosen to include most of the shock-heated hot gas in the galaxy around the virial temperature. 
    
    Figure~\ref{fig: hx-sx-ratio_maps} presents an image of the ratio of the projected surface brightness in hard X-ray versus the soft X-ray for a galaxy with an ID number 479938 from our sample. The distribution of the ratio roughly follows the morphology of the hot gas distribution shown in Figure~\ref{fig: density_maps}. The typical hard-to-soft X-ray ratio at the outskirts of the gas distribution is around $10^{-5}$-$10^{-4}$ and the corresponding temperature is close to the virial temperature of the distribution. In the central $\sim 50\,{\rm kpc}$ part of the distribution, the hard-to-soft X-ray ratio reaches around $10^{-3}$-$10^{-2}$.

\section{Shape Analysis}
\label{sec: shape_analysis}

    In this section, we present our method for computing the shape of the gas particles, the Local Shell Iterative Method (LSIM), and present the results of our analysis of the resultant shape parameters.

\subsection{The Local Shell Iterative Method (LSIM)}
\label{sec: LSIM}
    To compute the shape of the galactic gas in our galaxy sample, we employ the Local Shell Iterative Method (LSIM) as introduced in \citet{DMHaloMorph}. The shape parameters $s$ and $q$ are defined as follows:
\begin{equation} \label{eq: shape_param_sq}
s \equiv \frac{a}{c} ,~~ q \equiv \frac{b}{c}, 
\end{equation}
    where $a$, $b$, and $c$ refer to the semi-minor, intermediate, and semi-major axes of the ellipsoid, respectively. We begin by splitting the distribution into $100$ logarithmic, radially thin shells in the range between $2$ to $202 \si{\kilo\parsec}$. These shells are initially spherical with $a=b=c=r_{\rm sph}$, where $r_{\rm sph}$ refers to the mean spherical radius of the shell. We iteratively compute the reduced inertia tensor (Equation~\ref{eq: intertia_tensor}) and distort each initially spherical shell until the algorithm converges. The criteria for convergence are described later in this section. The reduced inertia tensor is defined below:
\begin{equation} \label{eq: intertia_tensor}
    I_{ij}(\leq r_{\rm sph}) \equiv \left(\frac{1}{M_{\rm gas}}\right)\sum_{n=1}^{N_{\rm part}} \frac{m_nx_{n,i}x_{n,j}}{R_n^2(r_{\rm sph})}, ~~ i,j = (1,2,3),
\end{equation}
    where $M_{\rm gas}$ is the total mass of gas particles inside the shell (with the axis lengths $a(r_{\rm sph})$, $b(r_{\rm sph})$, and $c(r_{\rm sph})$), $N_{\rm part}$ is the corresponding particle number, $m_n$ is the associated gas mass, $x_{n,i}$ is the $i$-th coordinate of the $n$th particle, and $R_n(r_{\rm sph})$ is the elliptical radius of the $n$th particle. $R_n(r_{\rm sph})$ is defined as:
\begin{equation} \label{eq: radius}
    R_n^2(r_{\rm sph}) \equiv \frac{x_n^2}{a^2(r_{\rm sph})}+\frac{y_n^2}{b^2(r_{\rm sph})}+\frac{z_n^2}{c^2(r_{\rm sph})}, 
\end{equation}
    where $x_n$, $y_n$, and $z_n$ are the Cartesian coordinates of the $n$th particle.
    
    As the axis lengths $a(r_{\rm sph})$, $b(r_{\rm sph})$, and $c(r_{\rm sph})$ are initially unknown, we compute each at a given radius $r_{\rm sph}$ by iteratively determining the reduced inertia tensor for all particles within that shell. We then diagonalize the reduced inertia tensor to obtain the eigenvalues and eigenvectors used to deform the initial sphere to an ellipsoid, while keeping the enclosed volume of the shells fixed. Using the eigenvalues of the reduced inertia tensor, we infer the axis lengths of the ellipsoid as follows: 
\begin{equation} \label{eq: abc}
\begin{split}
    a=\frac{r_{\rm sph}}{(abc)^{1/3}}\sqrt{\lambda_1}, \\ b=\frac{r_{\rm sph}}{(abc)^{1/3}}\sqrt{\lambda_2}, \\ c=\frac{r_{\rm sph}}{(abc)^{1/3}}\sqrt{\lambda_3},
\end{split}
\end{equation}
    where $\lambda_i$ ($i=1, 2, 3$) are the eigenvalues of the reduced inertia tensor.
    
    We then rotate the coordinates of all gas particles within the shells to the frame of the principal axes given by the eigenvectors of the reduced inertia tensor. We compute the shape parameters $s$ and $q$ for each iteration. In addition, the residuals of both $s$ and $q$ are computed at each iteration and the algorithm converges when the residuals drop to values below $\max[((s-s_{\rm old})/s)^2, \; ((q-q_{\rm old})/q)^2] \leq 10^{-3}$, where the 'max' refers to the greater value of the two aforementioned quantities and $s_{\rm old}$ and $q_{\rm old}$ are the shape parameters computed from the previous iteration. Furthermore, convergence requires that at least $10^3$ particles be present in each shell \citep{zemp}. Finally, we compute the triaxiality parameter 
    
\begin{equation} \label{eq: shape_param_T}   T \equiv \frac{1-q^2}{1-s^2},
\end{equation}
    where $s$ and $q$ are defined in Equation~\ref{eq: shape_param_sq}.

\begin{table}[h]
\centering
\caption{Median and 16th--84th percentiles for the shape parameters of each gas temperature regime, the gas as a whole, the stellar distributions, and the dark matter halos of all 25 MW-like galaxies in our sample.} \label{tab: median_percentile}
\begin{tabular}{lccc}
\hline
\hline
\multicolumn{1}{c}{} & $s$                         & $q$                         & $T$ \\ \hline \hline
Cold Gas             & ${0.315}_{-0.111}^{+0.107}$ & ${0.686}_{-0.089}^{+0.121}$ & ${0.619}_{-0.219}^{+0.122}$ \\ \hline
Warm Gas             & ${0.305}_{-0.084}^{+0.137}$ & ${0.697}_{-0.226}^{+0.169}$ & ${0.635}_{-0.295}^{+0.214}$ \\ \hline
Hot Gas              & ${0.499}_{-0.130}^{+0.099}$ & ${0.774}_{-0.101}^{+0.092}$ & ${0.570}_{-0.210}^{+0.155}$ \\ \hline
All Gas              & ${0.380}_{-0.111}^{+0.157}$ & ${0.752}_{-0.190}^{+0.133}$ & ${0.568}_{-0.259}^{+0.224}$ \\ \hline
Stellar Distribution & ${0.543}_{-0.134}^{+0.120}$ & ${0.913}_{-0.102}^{+0.044}$ & ${0.270}_{-0.155}^{+0.179}$ \\ \hline
Dark Matter          & ${0.739}_{-0.083}^{+0.061}$ & ${0.912}_{-0.046}^{+0.050}$ & ${0.333}_{-0.173}^{+0.246}$ \\ \hline
\end{tabular}
\end{table}

\subsection{Shape Analysis Results} \label{sec: analysis}

\begin{figure*}[hbt]
\centering
\includegraphics[width=0.8\paperwidth]{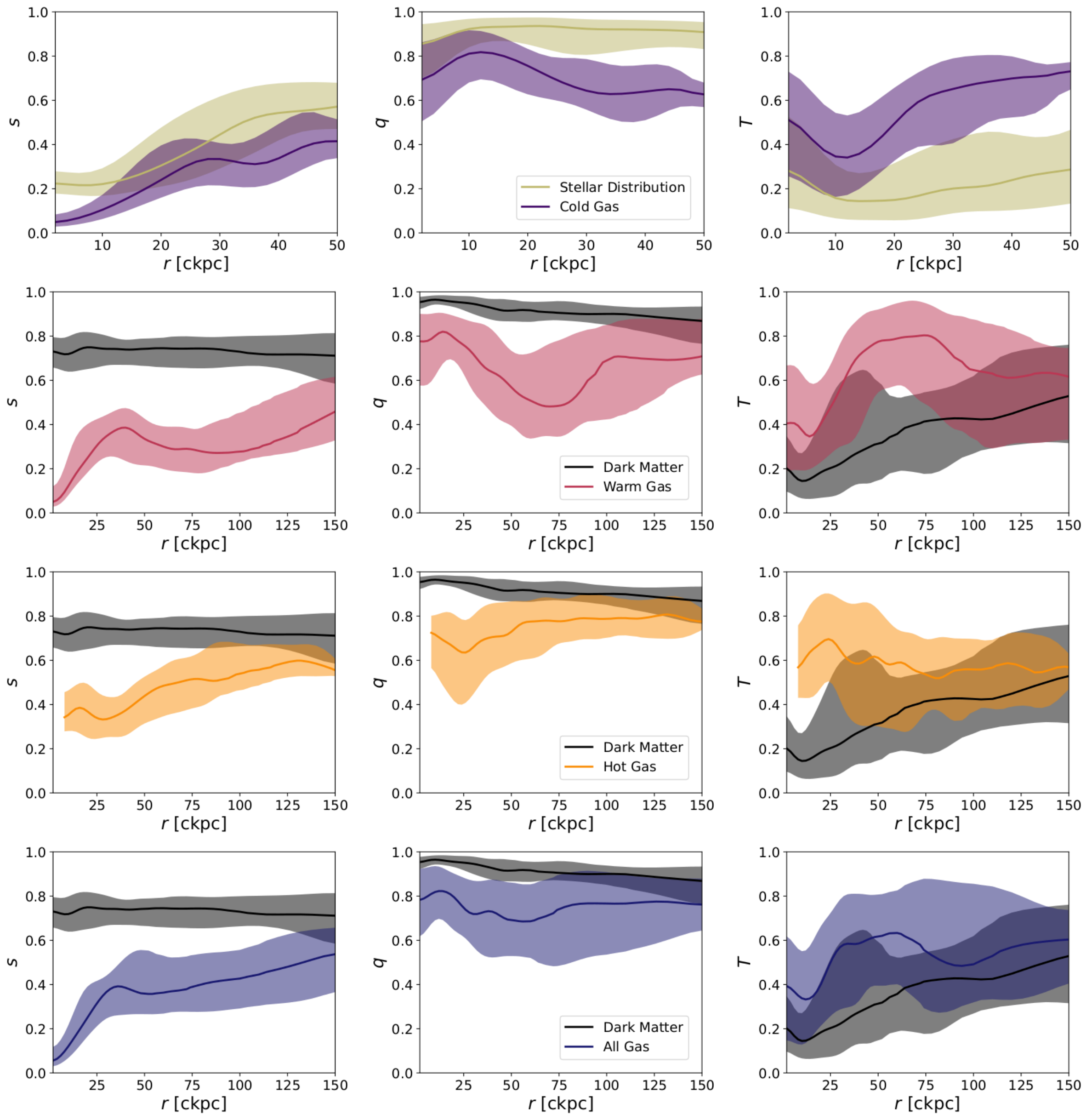}
\caption{Radial profiles of the median values, shown as solid lines, of the shape parameters $s$, $q$, and $T$ for all 25 MW-like galaxies in our sample as a function of the galactocentric radius with their 16th and 84th percentiles illustrated by the filled bands. Row 1 shows the radial profile of the median values of the shape parameters for the cold gas (purple) and the stellar distributions (yellow) \citep{StarMorph}. Rows 2-4 show the radial profiles of the median values of the shape parameters for the warm (red), hot (orange), and all gas (blue) distributions and the dark matter halos (black) \citep{DMHaloMorph}.
}
\label{fig: median_plot}
\end{figure*}

\begin{figure*}
\centering
\includegraphics[width = 0.8\paperwidth]{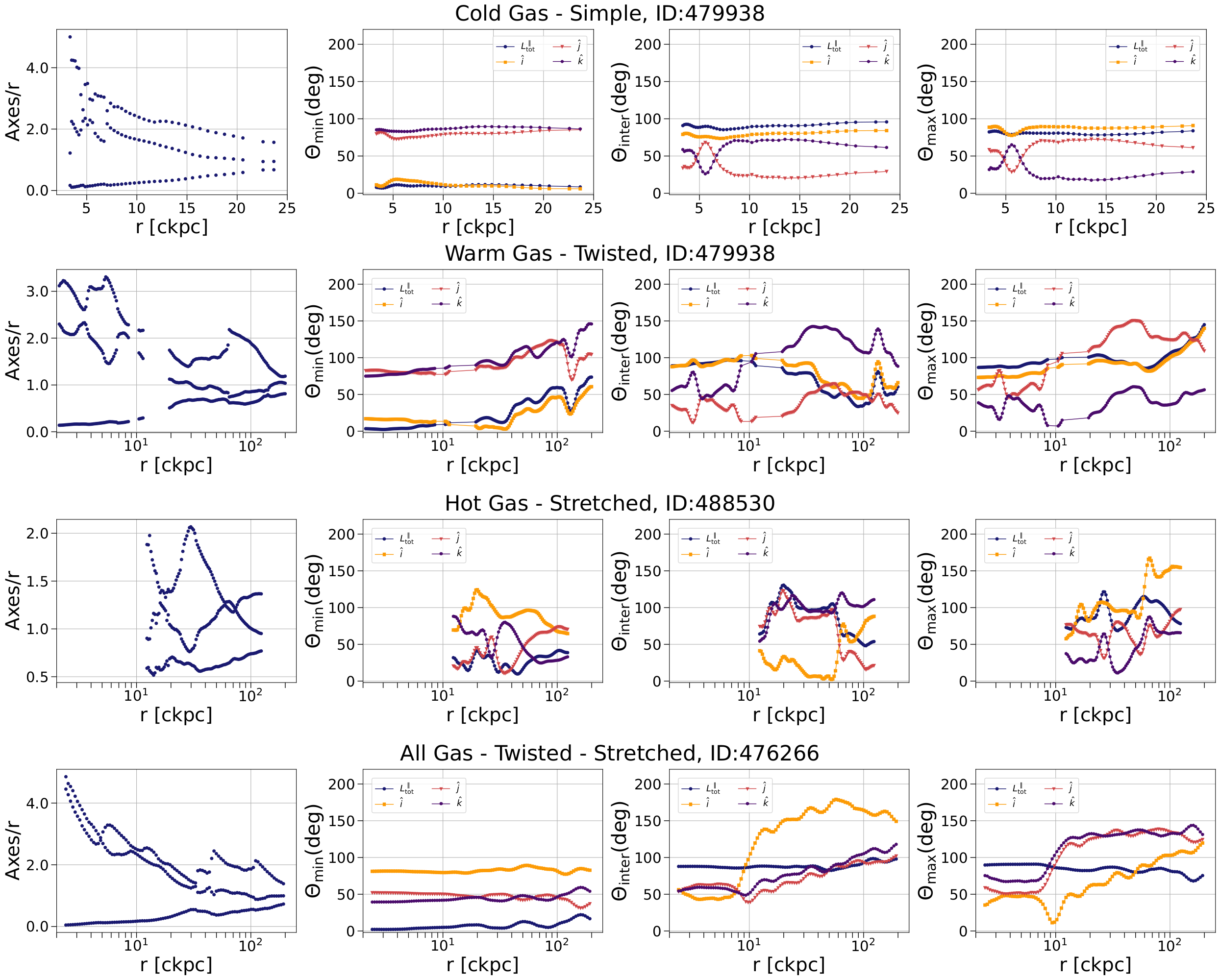}
\caption{Radial profiles of the axis lengths ($a$, $b$, and $c$) to radii ratios and the angles between the eigenvectors of the reduced inertia tensor and the total angular momentum of three galaxies within our sample. Row $1$ shows the radial profile of a simple cold gas distribution, row $2$ shows the radial profile of a twisted hot gas distribution, row $3$ shows the radial profile of a stretched hot gas distribution, and row $4$ shows the radial profile of a twisted-stretched distribution composed of all gas particles. The first panel in each row shows the ratios of each axis length to the radius as a function of galactocentric radius. Panels $2$-$4$ in each row show the angles for the minimum, intermediate, and maximum eigenvalues of the reduced inertia tensor to the $\uveci$, $\uvecj$, and \bm{$\hat{k}$} unit vectors as a function of galactocentric radius. Rows $1$ through $3$ are taken from various profiles showing the three major classification schemes: simple, twisted, and stretched in the cold, warm, and hot gas temperature regimes, respectively. Unlike the other gas regimes, the radial profile of hot gas begins at $10 \, \rm ckpc$ and ends at $\sim 150 \, \rm ckpc$. This is due to an insufficient number of particles at smaller and larger radii, so the LSIM algorithm does not converge. The final row shows a distribution that displays signals of both stretching and twisting in its radial profile of all gas particles.
\label{fig: eigen_angle}}
\end{figure*}

    Here, using the LSIM algorithm described in Section~\ref{sec: LSIM}, we compute the shape parameters for all of the galaxies in our sample. Table~\ref{tab: median_percentile} summarizes the median and $16$th($84$th) percentile for shape parameters. From top to bottom, we present the results for cold, warm, hot, and all gas, respectively. In addition, we computed and displayed the shape parameters for the stellar distributions and the DM halos with the LSIM algorithm for the sake of comparison (see Section~\ref{sec: comparison}). 
    
    After obtaining the median values of the triaxiality parameter, $T$, for the gaseous distributions, we use the criteria from \citet{DMHaloMorph} (summarized in Table~\ref{tab: ellipsoid_type}) to classify the shape of the gaseous distributions. From the median $T$ values in Table~\ref{tab: median_percentile}, we find that cold gas, warm gas, hot gas, and all gas distributions typically display triaxial ellipsoid morphologies.
    
    We also present the median values for each shape parameter $s$, $q$, and $T$ and their 16th and 84th percentiles as a function of the galactic radius in Figure~\ref{fig: median_plot}. From top to bottom, the radial profiles of these shape parameters are shown for cold, warm, hot gas, and all gas where the columns correspond to $s$, $q$, and $T$ from left to right. Also overlaid on these plots are the median and 16th and 84th percentiles of the shape parameters for the stellar distributions (yellow, top row) and the dark matter halos (black, remaining rows).

\section{Gaseous Distribution Shape Classification} \label{sec: shape_class}

    In the following sections, we analyze the shapes of the individual gaseous distributions in our sample of galaxies. Figure~\ref{fig: eigen_angle} presents the radial profiles of the axes to radii ratios, the angles between the eigenvectors\footnote{See Appendix~\ref{appendix:eigenvectors} for information concerning the orientation of these eigenvectors.} of the reduced inertia tensor (principal axes), and the total angular momentum with respect to the Cartesian unit vectors $\uveci$, $\uvecj$, and \bm{$\hat{k}$} for two galaxies in our sample. 

    Using the method described in Sections~\ref{sec: simple},~\ref{sec: stretched},~\ref{sec: twisted}, and~\ref{sec: twisted-stretched}, we implement these radial profiles in the classifications of the shape of the individual gaseous distributions for each temperature regime (summarized in Table~\ref{tab: tempclass}) into the following categories: simple, twisted, stretched, and twisted-stretched distributions (adapted from the classification scheme originally presented in \citet{StarMorph} for stellar distributions and \citet{DMHaloMorph} for dark matter halos). The results of these classifications are summarized in Table~\ref{tab: classifications}. 

    As was first addressed in \citet{DMHaloMorph}, LSIM produces slightly noisier radial profiles than other methods, such as EVIM, due to local fluctuations. These small-scale fluctuations can lead to additional crossings of the eigenvalues and artificially inflate the number of distributions classified as stretched (see Section~\ref{sec: stretched}). To prevent spurious results, we employ a Savitzky–Golay filter \citep{savgol} with SciPy\_python \citep{2007CSE.....9c..10O}, to smooth each radial profile.
    
    In Figure~\ref{fig: eigen_angle}, we show a collection of profiles that display all three of our major classification schemes: simple, twisted, and stretched, which correspond to cold, warm, and hot gas, respectively. For the gas as a whole, we present a distribution that displays signals of both twisting and stretching.

\begin{table}[t]
\centering
\caption{Types of ellipsoids as determined by their triaxiality parameter defined in Equation~\ref{eq: shape_param_T}. \label{tab: ellipsoid_type}}
\begin{tabular}{ll} 
\hline 
\hline
                   & Triaxiality Parameter             \\ \hline \hline
Oblate Spheroid    & $T = 0$ \\ \hline
Oblate Ellipsoid   & $T < 0.33$                        \\ \hline
Triaxial Ellipsoid & $0.33 \leq T \leq 0.66$           \\ \hline
Prolate Ellipsoid  & $T > 0.66$                        \\ \hline
Prolate Spheroid   & $T = 1$                           \\ \hline
\end{tabular}
\end{table}

\begin{table*}[h]
\centering
\caption{Summary of shape classifications for all 25 Milky Way-like galaxies for each temperature regime, stellar distributions \citep{StarMorph}, and dark matter \citep{DMHaloMorph} halos computed with the LSIM algorithm in the following categories: simple, twisted, stretched, and twisted-stretched distributions. Distributions labeled "not converged" are those distributions that do not converge in the LSIM algorithm (see Section~\ref{sec: LSIM} for LSIM convergence criteria).} \label{tab: classifications}
\centering
\begin{tabular}{lllllll}
\hline
\hline
\multicolumn{1}{l}{ID} & Cold Gas          & Warm Gas          & Hot Gas           & All Gas           & Stellar Distribution & Dark Matter       \\ \hline \hline
476266 & twisted           & twisted                                     & stretched                                   & twisted-stretched & twisted              & stretched         \\ \hline
478216 & not converged                & stretched         & twisted-stretched & simple            & twisted-stretched    & twisted           \\ \hline
479938 & simple            & twisted                                     & stretched                                   & twisted           & twisted              & stretched         \\ \hline
480802 & not converged                & twisted           & twisted-stretched                           & twisted-stretched & twisted-stretched    & twisted-stretched \\ \hline
485056 & stretched         & twisted           & twisted                                     & twisted           & twisted              & twisted-stretched \\ \hline
488530 & twisted           & twisted-stretched                           & stretched                                   & twisted-stretched & twisted              & twisted-stretched \\ \hline
494709 & twisted           & twisted-stretched & twisted-stretched & simple            & twisted-stretched    & twisted           \\ \hline
497557 & simple            & twisted-stretched                           & twisted                                     & twisted-stretched & twisted              & twisted           \\ \hline
501208 & twisted           & twisted-stretched & twisted           & twisted-stretched & twisted-stretched    & twisted-stretched \\ \hline
501725 & twisted-stretched & twisted                                     & twisted-stretched & twisted           & twisted              & twisted           \\ \hline
502995 & twisted           & twisted                                     & twisted-stretched & twisted           & twisted              & twisted           \\ \hline
503437 & twisted           & twisted-stretched                           & twisted-stretched & twisted           & twisted              & twisted-stretched \\ \hline
505586 & twisted           & stretched                                   & twisted-stretched                           & stretched         & twisted              & twisted-stretched \\ \hline
506720 & twisted           & twisted-stretched & twisted-stretched                           & twisted-stretched & twisted-stretched    & twisted-stretched \\ \hline
509091 & twisted           & stretched         & twisted           & twisted-stretched & twisted              & twisted-stretched \\ \hline
510585 & simple            & stretched         & twisted           & twisted           & twisted              & twisted-stretched \\ \hline
511303 & twisted           & twisted                                     & twisted                                     & twisted           & twisted-stretched    & twisted           \\ \hline
513845 & not converged                & twisted-stretched                           & twisted-stretched & twisted           & twisted-stretched    & stretched         \\ \hline
519311 & twisted           & twisted-stretched                           & twisted-stretched & twisted-stretched & twisted-stretched    & twisted           \\ \hline
522983 & twisted           & twisted                                     & twisted-stretched & twisted-stretched & twisted-stretched    & twisted           \\ \hline
523889 & not converged                & stretched         & twisted                                     & simple            & twisted              & twisted           \\ \hline
529365 & simple            & twisted-stretched & twisted                                     & twisted           & twisted-stretched    & twisted           \\ \hline
530330 & twisted           & twisted                                     & twisted                                     & twisted           & twisted-stretched    & twisted           \\ \hline
535410 & twisted           & twisted           & twisted                                     & stretched         & twisted-stretched    & twisted           \\ \hline
538905 & twisted-stretched & stretched                                   & twisted-stretched                           & twisted-stretched & twisted              & stretched      \\  \hline
\end{tabular}
\end{table*}

\subsection{Simple Distributions} \label{sec: simple}

    The first category that we consider is simple distributions. Simple distributions are those whose radial profiles display the eigenvector associated with the smallest eigenvalue that is parallel to the total angular momentum throughout the radial profile with little or no change in angle as a function of the radius. In addition, the remaining eigenvectors also show little-to-no directional changes. These profiles also typically display well-separated eigenvalues, whose ordering of their magnitude does not change with radius. Since the requirement of well-separated eigenvalues (which was originally introduced in \citet{DMHaloMorph}) will favor triaxial distributions, we also classify distributions whose eigenvalues are approximately equal (and may frequently cross), as would be the case in nearly spherical distributions, but show little-to-no rotation (after smoothing with the Savitzky–Golay filter) in their radial profiles as simple distributions. 
    
    In our sample of 25 MW-like galaxies, we find that only cold gaseous distributions and distributions composed of all gas have an entirely simple profile. Specifically, 4 cold gas distributions and 3 all gas distributions display simple radial profiles. An example of a simple distribution is presented in the first row of Figure~\ref{fig: eigen_angle}. We also show alternative visualizations of simple distributions in Figure~\ref{fig: simple_projection} and Figure~\ref{fig: simple_3d} in Appendix \ref{appendix:viz}. Galaxies part of the simple class, experience a very mild level of rotation in their radial profiles.

\subsection{Twisted Distributions} \label{sec: twisted}

    The next classification that we consider is twisted distributions. The members of this class display significant rotation in the radial profiles of their eigenvectors. More specifically, the minimum, intermediate, and maximum eigenvectors experience a gradual rotation of approximately 50-100 degrees throughout their radial profile with respect to the fixed $\uveci$, $\uvecj$, and \bm{$\hat{k}$} unit vectors and the total angular momentum. In some cases, twisted distributions display the eigenvalues approaching each other, yet the axes-to-radii ratios do not cross. However, as was noted in Section~\ref{sec: simple}, distributions with approximately equal eigenvalues, such as spherical distributions, may display frequent eigenvalue crossings. These distributions are still classified as twisted if they display gradual rotation in their radial profiles after smoothing with the Savitzky–Golay filter.
    
    A large percentage of distributions in our sample exhibit purely twisted radial profiles. Our sample of purely twisted distributions consists of 14 cold gas distributions, 10 warm distributions, 10 hot gas distributions, and 10 all gas distributions. An example of a twisted distribution is shown in the second row of Figure~\ref{fig: eigen_angle}. We also show alternative visualizations of twisted distributions in Figure~\ref{fig: twisted_projection} and Figure~\ref{fig: twisted_3d} in Appendix~\ref{appendix:viz}. 

\subsection{Stretched Distributions} \label{sec: stretched}

    The next category of distributions that we present is the stretched class defined as when two eigenvalues of the reduced inertia tensor approach each other with a reversal of their orders. Since their corresponding eigenvectors are orthogonal, the stretching of their axis lengths is seen in terms of an abrupt rotation of approximately $90^\circ$ in their eigenvectors. As was addressed in Section~\ref{sec: shape_class}, the frequent crossing of eigenvalues due to two approximately equal axes could introduce false classifications of stretched distributions in oblate, prolate, nearly spherical, or disk-like morphologies. For these classifications, we consider only those distributions with significant eigenvalue crossings accompanied by clear, abrupt $\sim 90^\circ$ rotations after smoothing with the Savitzky–Golay filter. One limitation of this classification scheme is distributions that have axes that stretch, but their eigenvalues do not cross. With our current classification scheme, these distributions will be either classified as simple or twisted, depending on the change in orientation of their eigenvectors. However, for the purposes of this work, we consider stretched distributions as only those with crossings of eigenvalues that result in abrupt $\sim 90^\circ$ rotations.
    
    In our sample, we have 1 cold gas distribution, 6 warm gas distributions, 3 hot gas distributions, and 2 all gas distributions in this class. An example of a stretched distribution is shown in the third row of Figure~\ref{fig: eigen_angle}. We also show alternative visualizations of stretched distributions in Figure~\ref{fig: stretched_projection} and Figure~\ref{fig: stretched_3d} in Appendix~\ref{appendix:viz}.

\subsection{Twisted-stretched Distributions} \label{sec: twisted-stretched}

    The final classification we group our gaseous distributions into is the twisted-stretched category, which displays both distinct stretching and twisting in their radial profiles. As defined above, the stretched portions of these distributions show a crossing in the eigenvalue profiles as one principal axis stretches in magnitude and exceeds the magnitude of another which as mentioned above is accompanied by an approximately $90^\circ$ rapid rotation in their associated eigenvectors. In addition, these distributions also experience some level of gradual rotations in their aforementioned eigenvectors.
    
    In our sample of 25 MW-like galaxies, 2 cold gas distributions, 9 warm gas distributions, 12 hot gas distributions, and 10 of all gas distributions are part of the twisted-stretched class. An example of a twisted-stretched distribution is shown in the last row of Figure~\ref{fig: eigen_angle}.

\section{Comparisons} \label{sec: comparison}

    Here, we present quantitative and qualitative comparisons between the following items: gaseous distributions in our individual temperature regimes (Section~\ref{sec: gas_comparison}), cold gas distributions and stellar distributions (Section~\ref{sec: star_comparison}), and warm, hot, and all gas distributions to dark matter halos (Section~\ref{sec: DM_comparison}). The dark matter halos and stellar distributions are described in more detail in \citet{DMHaloMorph} and \citet{StarMorph}, respectively.

\subsection{Gas Temperature Regime Comparison} \label{sec: gas_comparison}

    Here we present a comparison between gaseous distributions in the cold, warm, hot, and all gas temperature demarcations, as defined in Section~\ref{sec: temp} Table~\ref{tab: tempclass}. The first set of comparisons that we consider is the gaseous distribution morphology classifications presented in Section~\ref{sec: shape_class}. The morphology classifications for each galaxy in our sample of 25 MW-like galaxies for each temperature regime (see Table~\ref{tab: tempclass}) are summarized in Table~\ref{tab: classifications}.
    
    We find that the vast majority ($77\%$) of all distributions show twisting at some point in their radial profiles. Specifically, $64\%$ of cold gas distributions, $76\%$ of warm gas distributions, $88\%$ of hot gas distributions, and $80\%$ of all gas distributions show twisting in their radial profiles and are classified as either twisted or twisted-stretched distributions. Of these distributions, nearly half ($44\%$) experience purely twisted radial profiles. Interestingly, all gas distributions experience the lowest number of purely twisted distributions (40\%), while cold gas has the most purely twisted distributions (56\%).
    
    Furthermore, we conclude that gaseous distributions tend to become more stretched at higher temperatures. Only 12\% of cold gas distributions experience stretching at some point, while warm and hot gas experience some form of stretching in $60\%$ of their distributions. Furthermore, when we look at the gas as a whole, nearly half ($45\%$) of distributions experience stretching at some point in their radial profiles. This high rate of twisting in the cold gas is likely influenced by the rotation of the disk, while the more extended warm and hot gas is less affected by this rotation and, thus, exhibits more stretching in their radial profiles.

    Considering purely stretched and simple distributions for all galaxies in each temperature regime, we conclude that the minority of distributions exhibit these features.  Specifically, only 12\% of gaseous distributions are classified as purely stretched. Specifically, only 4\% of cold gas, $24\%$ of warm gas, $12\%$ of hot gas, and $8\%$ of all gas distributions are purely stretched. The least common distribution type in our sample is simple distributions. Only $7\%$ of distributions are classified as simple. These distributions appear only in cold gas and all gas temperature regimes at $16\%$ and $12\%$ respectively.

    Next, we present a comparison between the individual temperature regimes by examining the shape parameters $s$, $q$, and $T$, with a primary focus on the triaxiality parameter $T$ (Equation~\ref{eq: shape_param_T}). For this comparison, we employ Table~\ref{tab: median_percentile} and Figure~\ref{fig: median_plot}.
    
    Using Table~\ref{tab: median_percentile}, we find that all temperature regimes show similar values for the triaxiality parameter $T$ with median values of ${0.619}$, ${0.635}$, ${0.570}$, ${0.568}$ for cold gas, warm gas, hot gas, and all gas, respectively. Using the median values and the types of ellipsoids as defined in Table~\ref{tab: ellipsoid_type}, we find that cold gas, warm gas, hot gas, and all gas distributions show median values of the triaxiality parameter $T$ that correspond to triaxial ellipsoids.
    
    The radial profiles of the median values of the shape parameters (shown in Figure~\ref{fig: median_plot}) for each temperature regime are also remarkably similar. Perhaps the most striking similarity is that the radial profiles of warm, hot, and all gas distributions for the triaxiality parameter $T$ all begin to align with $T\sim0.6$ at $r\sim150 \rm kpc$. However, the warm gaseous distribution radial profile shows the largest deviation from the norm. When considering the radial profile of $T$ for warm gas, we find that the warm gas is incredibly prolate in the region between approximately $50 \rm kpc$ to $125 \rm kpc$. Cold gas, to a lesser degree, also appears to be prolate between approximately $45 \rm kpc$ and $60 \rm kpc$ and traces a profile similar to that of the warm gas. However, the radial scales between these two temperature regimes are not directly comparable. Another interesting aspect is that the radial profile of $T$ for hot gas appears to be, in general, the approximate inverse of the other gas temperature regime radial profiles. This is likely due to the lack of hot gas in the inner $\sim10 \rm kpc$ of the disk, which is evident in Figure~\ref{fig: median_plot}. The hot gas profile does not display the initial decrease in $T$ from the center to the edge of the disk. This is particularly evident in the cold gas radial profile which shows a significant decrease in $T$ in the inner $\sim15 \rm kpc$.

\subsection{Cold Gas to Stellar Distribution Comparison} \label{sec: star_comparison}

    Next, we present a comparison of the cold gas distributions to the stellar distributions. Again, we consider Tables~\ref{tab: median_percentile} and~\ref{tab: classifications} and Figure~\ref{fig: median_plot} in our comparisons. We adopt the morphology classifications for the stellar distribution from \citet{StarMorph} (included for reference in Table~\ref{tab: classifications}). \citet{StarMorph} also employs the LSIM algorithm (see Section~\ref{sec: LSIM}). We use data that was generated in their shape calculations to determine the median values and 16th (84th) percentiles utilized in Table~\ref{tab: median_percentile} and Figure~\ref{fig: median_plot}.
    
    Considering the gaseous distribution shape classifications (summarized in Table~\ref{tab: classifications}), the majority ($56\%$) of cold gas distributions are purely twisted. Interestingly, the majority of stellar distributions ($52\%$) presented in \citet{StarMorph} are also purely twisted. However, when including twisted-stretched distributions, we find that $64\%$ of cold gas distributions and $100\%$ of stellar distributions show twisting at some point in their radial profiles. Therefore, most galaxies tend to show twisting in both the cold gas distributions and the stellar distributions. This is not particularly surprising considering that our sample of galaxies is MW-like and shows significant rotation in the disks. Furthermore, cold gas is closely coupled to the star-forming regions within the disk. However, we do observe some deviation from the stellar distribution in the cold gas distribution morphologies with $16\%$ of distributions being classified as simple and $4\%$ of distributions being classified as purely stretched. It should be noted though that $16\%$ of cold gas distributions do not converge with the LSIM algorithm.
    
    Next, considering a direct distribution-to-distribution comparison using Table~\ref{tab: classifications} and \citet{StarMorph}, we find that $61.5\%$ of the galaxies are classified as purely stretched in both cold gas and stellar distributions. Interestingly, zero galaxies are classified as twisted-stretched in both cold gas and stellar distributions, but $66.7\%$ of twisted-stretched stellar distributions are purely twisted in the cold gas distributions. Very few cold gas distributions deviate from the morphology of the stellar distributions completely, with $8.3\%$ twisted-stretched stellar distributions being classified as simple in cold gas and $7.7\%$ purely twisted stellar distributions being classified as simple in cold gas distributions. The remaining deviation is the non-convergent cold gas distributions. 
    
    Finally, by examining Table~\ref{tab: median_percentile} and Figure~\ref{fig: median_plot}, there seems to be a correlation between the morphological properties of stellar distributions and cold gas. Remarkably, cold gas displays a significantly more prolate morphology than the stellar distribution with median triaxiality parameters of $T=0.619$ and $T=0.270$ respectively. While a majority of cold gas resides in the disk (as is apparent from Figure~\ref{fig: density_maps}), there is a significant amount of cold gas that drives the overall morphology to be much more prolate than the disky, oblate stellar distributions. This is very apparent when we examine the first row of Figure~\ref{fig: median_plot}. Although the cold gas has a radial profile similar to the stellar distribution, the median $T$ of the cold gas is significantly higher than the median $T$ for the stellar distribution throughout the entire radial profile. However, these values do begin to slightly align with $T \sim 0.4$ at $r \sim 90 \rm kpc$. In addition, the decrease in the median value of $T$ from $r=0 \rm kpc$ to $r \sim 15 \rm kpc$ in the cold gas radial profile is almost entirely parallel to the stellar distribution, suggesting that the cold gas traces the central bulge of the stars nearly exactly, but with a significantly more prolate morphology.

\subsection{Gas to Dark Matter Comparison} \label{sec: DM_comparison}

    In this section, we compare the morphology classifications of the gaseous distributions in each temperature regime to those of the dark matter halos. As \citet{DMHaloMorph} uses the Enclosed Volume Iterative method as their main algorithm to determine the morphology of the dark matter halos, we use the LSIM algorithm to compute the radial profiles and classify each dark matter halo using our classification scheme (as outlined in Sections~\ref{sec: shape_analysis} and~\ref{sec: shape_class}). From the LSIM dark matter radial profiles, we determine that there are 0 simple, 12 twisted, 4 stretched, and 9 twisted-stretched dark matter halos. We summarize these classifications in Table~\ref{tab: classifications}. In addition, we computed the median and 16th (84th) percentiles with data generated from the LSIM algorithm.
    
    Upon direct comparison of the individual gaseous distributions to the dark matter halos classified with the LSIM algorithm using Table~\ref{tab: classifications}, we find a strong correlation between the gas and dark matter morphology classifications with increasing gas temperature. For purely twisted gaseous distributions, we find that $50\%$ of cold gas, $60\%$ of warm gas and hot gas distributions are also twisted in dark matter. For purely stretched gaseous distributions, $0\%$ of cold gas, $16.7\%$ of warm gas, and $66.7\%$ of hot gas distributions are stretched in dark matter. Twisted-stretched distributions also display this correlation with $0\%$ of cold gas, $44.4\%$ of warm gas, and $33.3\%$ of hot gas distributions sharing the same classification as the dark matter halos. However, gaseous distributions that contain twisting at some point in their radial profile correlate to dark matter halos that show twisting at $87.5\%$, $84.2\%$, and $90.9\%$ for cold, warm, and hot gas respectively. Similarly, for stretching at some point in the radial profiles, $66.7\%$, $66.7\%$, and $46.7\%$ of cold, warm, and hot gas also show stretching in the dark matter halos. When considering gaseous distributions composed of all gas particles, we find that $50\%$ of twisted all gas distributions are twisted in dark matter, $0\%$ of stretched all gas distributions are stretched in dark matter, and $50\%$ of twisted-stretched all gas distributions are twisted-stretched in dark matter. Again, when considering twisting or stretching at some point in the radial profiles, $80\%$ of all gas distributions and dark matter show twisting, and $72.7\%$ of all gas distributions and dark matter halos show stretching.

    Considering the triaxiality parameter $T$ presented in Table~\ref{tab: median_percentile} and Figure~\ref{fig: median_plot}, we find that the gas of all temperature regimes, and as a whole, is significantly more prolate than the dark matter halos. Table~\ref{tab: median_percentile} shows a median $T$ for gaseous distributions of ${0.619}$, ${0.635}$, ${0.570}$, and ${0.568}$ for the cold, warm, hot, and all gas regimes, respectively. With an average value of $0.609$, the gaseous distributions have a median $T$ $2.8$ greater than the median $T$ for the dark matter halos. 

    The comparison between warm, hot, and all gas to dark matter halos is also presented in Figure~\ref{fig: median_plot} where the radial profiles of the median values (and their 16th and 84th percentiles) of the shape parameters $s$, $q$, and $T$ for all 25 galaxies are presented. While we have also compared cold gas to dark matter, we omit cold gas from the following due to the difference in the radial extent of cold gas and dark matter. The most striking difference in these profiles is in the first two columns, showing the shape parameters $s$ and $q$, where the gas radial profiles deviate significantly from the dark matter radial profiles. The final column, showing the radial profile of the triaxiality parameter $T$, shows lesser of a degree of deviation from dark matter. Here, the higher degree of prolate morphologies, as illustrated by Table~\ref{tab: median_percentile}, is also apparent. While warm, hot, and all gas show radial profiles of $T$ that are higher on average, warm gas shows the most extreme deviation from the dark matter $T$ radial profile.
    
    However, all of the shape parameter radial profiles align for dark matter at larger radii. As the temperature of the gas increases, the alignment of the median values and 16th (84th) percentiles happen at smaller radii, with this alignment beginning at roughly $150 \rm kpc$ and $100 \rm kpc$ for $s$, $125 \rm kpc$ and $45 \rm kpc$ for $q$, for warm and hot gas respectively. When considering the 16th (84th) percentile for the triaxiality parameter, there is significantly more overlap between hot gas and dark matter than between warm gas and dark matter. Hot gas and dark matter overlap for almost the entirety of the radial profile. Additionally, the median value of the triaxiality parameter for hot gas and dark matter have remarkably similar values after about $75 \rm kpc$. In contrast, the warm gas and dark matter triaxiality parameters are initially within 0.2 of each other before beginning to deviate from one another around $25 \rm kpc$, only until beginning to align again after about $60 \rm kpc$. 

\section{Connection to Observations and other Simulations} \label{sec: observations}

    Thus far, we have presented a theoretical framework of gas morphology with the IllustrisTNG TNG50 simulation using the LSIM algorithm to compute quantitative shapes of gas distributions in the cold, warm, and hot regimes. In this section, we connect these results to current and future observational constraints of galactic gas and the Circumgalactic Medium (CGM) and the work done with other hydrodynamical simulations. 
    
    To study the CGM, there are typically five main methodologies: absorption line studies of background sources, spectroscopic stacking, ``down-the-barrel'' spectroscopy, emission line maps, and hydrodynamic simulations. These four observational methodologies each have significant challenges and advantages. 

    Absorption line studies are perhaps the most sensitive method to extremely low column densities of the CGM ($N \simeq 10^{12} \, \rm{cm^{-2}}$) and offer the advantage of probing a wide range of densities while remaining invariant to redshift and host galaxy luminosity \citep{HB2022}. This method typically allows only one line of sight through the host CGM at a random galaxy radius, however, so comparisons to hydrodynamic simulations can only be done after averaging \citep{smith2019}. Spectroscopic stacking and ”down-the-barrel” have similar limitations. 
    
    Emission line maps, however, offer the ability to get large-scale maps of CGM gas through direct emission, making this the ideal method for comparison with our results from the TNG50 simulation. This method is, however, one of the most challenging methods mentioned thus far. The emission measure scales as $n^2$ and $n_H$ in the CGM is typically about $10^{-2}$ or less, making detecting this emission quite difficult. Some collaborations, such as \citet{Emission_map1}, have applied this methodology with the H\,\footnotesize I\normalsize-21 cm line to external galaxies, but their results are limited to the inner $\sim 10-20 \, \rm kpc$, significantly less than the spatial extent of the cold gas ($\sim 100 \, \rm ckpc$) we find in our work. \citet{21cmwarpeddisk} also generated an H\,\footnotesize I\normalsize-21 cm map of NGC 3178, finding evidence of warping and twisting in the extended neutral hydrogen disk out to $\sim 35 \, \rm kpc$. H\,\footnotesize I\normalsize-21 cm maps are also useful in probing DM halo properties. \citet{m3321cm} used H\,\footnotesize I\normalsize-21 cm emission to a radius of $\sim 25 \, \rm kpc$ to compute rotation curves of M33 and found that the dynamically relevant baryons at large radii are in gas form and the DM halo has a mass of $M=4.3(\pm1.0)\times10^{11} \, \rm M_\odot$, assuming a Navarro-Frenk-White (NFW) profile and $\rm \Lambda CDM$ cosmology.

    In this work, we find a strong correlation between hot gas distributions and DM halos. Therefore, it is of particular interest to study the hot CGM with soft X-ray (ideal for gas with $T \gtrsim 10^6 \rm K$) observations. However, the hot, diffuse CGM typically has very low surface brightness so observations with soft X-ray telescopes are challenging and require long exposure times. Nonetheless, several collaborations, such as \citet{xray2} and \citet{xray1}, have used observatories such as \textit{Chandra}, \textit{Suzaku}, and \textit{XMM-Newton} to observe individual hot gas halos. Like H\,\footnotesize I\normalsize-21 cm observations, however, gas properties are reliably measured to a smaller spatial extent ($\sim80-100 \, \rm kpc$) with these X-ray observations than we find in our work for hot gas distributions ($\sim 200 \, \rm kpc$). Despite the current challenges of X-ray emission mapping, future X-ray missions have promising potential to probe the extremely diffuse CGM gas at very large spatial extents. The \textit{Line Emission Mapper} (\textit{LEM}) X-ray probe, in particular, is a promising mission that will study the hot, diffuse CGM at scales well beyond the virial radii of galaxies \citep{LEM}.

    Hydrodynamical simulations offer the most detailed and controlled studies of the gaseous distributions of galaxies and are essential in a plethora of studies. These simulations, however, rely on current knowledge of the physical processes that shape galaxy formation and evolution, gas and DM physics, stellar and AGN feedback, star formation and supernovae, metal mixing and transport, cosmology, and many other processes, some of which are unresolved \citep{unresolved, CGM_review}. Even with these uncertainties, the knowledge gleaned from hydrodynamic simulations is powerful, especially when combined with observational data.

    There are several high-resolution hydrodynamical simulations currently being used to probe galaxy morphology and other properties including AURIGA \citep{auriga1, auriga2, auriga3}, EAGLE \citep{eagle1, eagle2, eagle3, eagle4}, FIRE \citep{fire1, fire2, fire3, fire4, fire5}, NIHAO-UHD \citep{nihao1, nihao2}, and IllustrisTNG \citep{tng1, tng2, tng3, tng4, tng5, tng6, tng7}. With such a wide variety of hydrodynamical simulations, we can deeply explore galaxy properties and produce realistic, reliable data, allowing for robust comparisons between simulations and observational data. For a more comprehensive overview of the methodologies for observations and hydrodynamic simulations mentioned in this section, and the CGM in general, see \citet{CGM_review}.

\section{Summary and Conclusions} \label{sec: conclusions}

    Here, we presented a detailed analysis of the morphology of gaseous distributions from 25 Milky Way-like galaxies from the IllustrisTNG TNG50 simulation and how this work connects to past, current, and future observations and other works done with hydrodynamic simulations. We selected galaxies with mass ranges of $(1-1.6)\times 10^{12} M_\odot$ that have more than $40\%$ of their stars rotationally supported.

    We inferred the temperature of the gas particles and grouped them into three temperature regimes: cold, warm, and hot gas (see Table~\ref{tab: tempclass}). We computed the hard-to-soft X-ray luminosity ratios for gas with temperatures above $0.5T_{\rm vir}$. We found that the hard-to-soft X-ray ratio at the outer portion of the hot gas distribution ranges from $10^{-5}$ to $10^{-4}$ and from $10^{-3}$ to $10^{-2}$ in the inner $\sim 50 \rm kpc$ of the distribution. 

    Next, we employed the Local Shell Iterative Method (LSIM) algorithm to compute the shape of the gas particles in our sample. The LSIM algorithm iteratively computes the reduced inertia tensor (Equation~\ref{eq: intertia_tensor}) for all particles within $100$ (initially spherical) logarithmic, radially thin shells. We then diagonalize the reduced inertia tensor to obtain the eigenvalues and eigenvectors. Finally, we employ the eigenvalues and eigenvectors to compute the aforementioned shape parameters and to create radial profiles of the axes/r ratios, the angles between the principal axes and the Cartesian unit vectors (and the total angular momentum), and the shape parameters $s$ and $q$. 

    We inferred the shape of gas particles in four different regimes; cold, warm, hot, and all gas. We then classified the gas particles in each of these into three distinct categories: simple, twisted, and stretched. Simple galaxies are those that display radial profiles with well-separated eigenvalues and little to no change in the orientation of the principal axes with respect to the Cartesian unit vectors. Galaxies part of the twisted class also show well-separated eigenvalues. In addition, they show a significant, but gradual rotation throughout their radial profiles. Galaxies part of the stretched class display a crossing of eigenvalues in their radial profiles, and thus, due to the orthogonality of the eigenvectors, show an abrupt $\sim 90^{\circ}$ rotation of the principal axes with respect to the Cartesian unit vectors. As many gaseous distributions in our sample show signals of both stretching and twisting, we group them into the twisted-stretched category. 

    In our sample of 25 MW-like galaxies, we classify 4 cold gas and 3 all gas as simple. No other temperature regimes display simple profiles. Stretched galaxies appear in every temperature regime with 1 cold gas distribution,6 warm gas distributions, 3 hot gas distributions, and 2 all gas distributions showing purely stretched radial profiles. Twisted cases occupy a much more significant percentage of gaseous distributions with 14 cold gas, 10 warm gas, 10 hot gas distributions, and 10 all gas distributions. The final category, twisted-stretched distributions, occupies 2 cold gas distributions, 9 warm gas distributions, 12 hot gas distributions, and 10 all gas distributions. In addition to computing the radial profiles of the shape parameters, we compute the median values and 16th (84th) percentiles of the shape parameters both as a function of galactocentric radius and as a median over all distributions in our sample. The resultant values for cold gas are as follows: $s={0.315}$, $q={0.686}$, and $T={0.619}$. The median values for warm gas are $s={305}$, $q={697}$, and $T={635}$. The median values for hot gas are $s={0.499}$, $q={0.774}$, and $T={0.570}$. Finally, the median values for all gas are as follows: $s={0.380}$, ${q=0.752}$, and $T={0.568}$. To make a robust comparison to previous works in this series (\citet{StarMorph} and \citet{DMHaloMorph}), we compute these values for the stellar distributions and dark matter halos as well. For the stellar distributions, we compute $s={0.543}$, $q={0.913}$, and $T={0.270}$. Finally, for dark matter, these values are $s={0.739}$, $q={0.912}$, and $T={0.333}$. From the aforementioned values, we determined that cold gas, warm gas, hot gas, and all gas distributions display primarily triaxial ellipsoid morphologies. 

    Following the quantification of the morphology of the gaseous distributions in our sample, we present detailed comparisons between gaseous distributions in our individual temperature regimes, comparing cold gas to stellar distributions, and gaseous distributions in each temperature regime with the dark matter halos. In the gas-to-gas comparisons, we find that the radial profiles of the median and 16th (84th) percentiles of the shape parameters are remarkably similar across temperature regimes, with the largest deviation being in warm gas due to the nearly prolate ellipsoid morphology. 
    
    In the cold gas to stellar distribution comparison, we find a strong correlation between gas and the stellar distribution morphology. Considering that the cold gas is closely coupled to the disk, this is not particularly surprising. Interestingly, though, cold gas shows a much more prolate morphology than the stellar distribution with median $T$ values of 0.619 and 0.270, respectively. This is also highly apparent in the radial profile of the median and 16th (84th) percentiles of the cold gas as the radial profile of $T$ for the cold gas resides above the radial profile of $T$ for the stellar distribution throughout the profile.

    The most interesting set of comparisons, however, is the gaseous distributions to the dark matter halos comparisons. As the previous morphology classification for dark matter from \citet{DMHaloMorph} employs the Enclosed Volume Iterative Method (EVIM), we reclassify these halos with radial profiles generated from the LSIM algorithm in order to present an accurate comparison. Again, we find a strong correlation between the gas morphology and the dark matter halo morphology. Remarkably, this correlation gets stronger with increasing gas temperature. For example, when considering twisted distributions, we find that 28\% of cold gas, 60\% of warm gas and hot gas distributions are twisted in both gas and dark matter. This correlation holds for the remaining temperature regimes (see Section~\ref{sec: DM_comparison}).

    In addition, the median value of the triaxiality parameter $T$ for the gaseous distributions compared to the dark matter halos may shed light on a very physically interesting situation. Considering the average of the median values of $T$ for each temperature regime presented above, we find that gas as a whole is significantly more prolate than dark matter. Specifically, the average value of the median $T$ for gas is $0.609$ while the median $T$ for dark matter is $0.333$. Since this value for gas is nearly a factor of two higher than that of dark matter, we hypothesize that the morphology of the gas is highly influenced by stellar and AGN feedback. Since the gas ejected in both stellar and AGN feedback is typically multi-phase \citep{refId0} and preferentially oriented perpendicular to the disk \citep{10.1093/mnras/sty229}, these phenomena could be driving gaseous distributions of all temperature regimes into more prolate cases. We hope to explore this hypothesis in a future publication, both in the IllustrisTNG TNG50 simulation and observationally.

\section*{Acknowledgements} \label{sec: acknowledgements}
    The SAO REU program is funded in part by the National Science Foundation REU and Department of Defense ASSURE programs under NSF Grants no.\ AST 1852268 and 2050813, and by the Smithsonian Institution. We also thank Matthew Ashby, Shep Doeleman, Yakov Faerman, Jonathan McDowell, and Jessica Werk for insightful conversations and invaluable comments. The authors thank the referee for their feedback, which greatly improved the quality of this manuscript. RE acknowledges the support from the Institute for Theory and Computation at the Center for Astrophysics as well as grant numbers NASA ATP 21-atp21-0077, NSF AST-1816420 and HST-GO-16173.001-A for very generous support. The authors thank the supercomputer facility at Harvard, where most of the simulation work was done. MV acknowledges support through an MIT RSC award, a Kavli Research Investment Fund, NASA ATP grant NNX17AG29G, and NSF grants AST-1814053, AST-1814259, and AST-1909831. The TNG50 simulation was realized with compute time granted by the Gauss Center for Supercomputing (GCS) under GCS Large-Scale Projects GCS-DWAR on the GCS share of the supercomputer Hazel Hen at the High-Performance Computing Center Stuttgart (HLRS). 

\software{
          illustris\_python \citep{TNGDataRelease, TNGFirstResultsFB, TNGFirstResultsEV},
          SciPy\_python \citep{2007CSE.....9c..10O},
          Matplotlib \citep{matplotlib},
          Numpy \citep{numpy}, 
          Py-SPHViewer \citep{pysphviewer}, 
          Python Programming Language \citep{python}, and
          TNG50 \citep{TNGDataRelease, TNGFirstResultsFB, TNGFirstResultsEV}
          }

\bibliography{references}

\begin{thebibliography}{}
\expandafter\ifx\csname natexlab\endcsname\relax\def\natexlab#1{#1}\fi
\providecommand{\url}[1]{\href{#1}{#1}}
\providecommand{\dodoi}[1]{doi:~\href{http://doi.org/#1}{\nolinkurl{#1}}}
\providecommand{\doeprint}[1]{\href{http://ascl.net/#1}{\nolinkurl{http://ascl.net/#1}}}
\providecommand{\doarXiv}[1]{\href{https://arxiv.org/abs/#1}{\nolinkurl{https://arxiv.org/abs/#1}}}

\bibitem[{{Allen} {et~al.}(2001){Allen}, {Schmidt}, \& {Fabian}}]{allen}
{Allen}, S.~W., {Schmidt}, R.~W., \& {Fabian}, A.~C. 2001, \mnras, 328, L37,
  \dodoi{10.1046/j.1365-8711.2001.05079.x}

\bibitem[{{Anders} \& {Grevesse}(1989)}]{AG1989}
{Anders}, E., \& {Grevesse}, N. 1989, \gca, 53, 197,
  \dodoi{10.1016/0016-7037(89)90286-X}

\bibitem[{{Anderson} {et~al.}(2016){Anderson}, {Churazov}, \&
  {Bregman}}]{xray2}
{Anderson}, M.~E., {Churazov}, E., \& {Bregman}, J.~N. 2016, \mnras, 455, 227,
  \dodoi{10.1093/mnras/stv2314}

\bibitem[{{Appleby} {et~al.}(2021){Appleby}, {Dav{\'e}}, {Sorini},
  {Storey-Fisher}, \& {Smith}}]{warm_tvir}
{Appleby}, S., {Dav{\'e}}, R., {Sorini}, D., {Storey-Fisher}, K., \& {Smith},
  B. 2021, \mnras, 507, 2383, \dodoi{10.1093/mnras/stab2310}

\bibitem[{Benitez-Llambay(2015)}]{pysphviewer}
Benitez-Llambay, A. 2015, py-sphviewer: Py-SPHViewer v1.0.0,
  \dodoi{10.5281/zenodo.21703}

\bibitem[{{Bertola} \& {Galletta}(1979)}]{bertola}
{Bertola}, F., \& {Galletta}, G. 1979, \aap, 77, 363

\bibitem[{{Bhattacharyya} {et~al.}(2023){Bhattacharyya}, {Das}, {Gupta},
  {Mathur}, \& {Krongold}}]{warmhot_MW}
{Bhattacharyya}, J., {Das}, S., {Gupta}, A., {Mathur}, S., \& {Krongold}, Y.
  2023, \apj, 952, 41, \dodoi{10.3847/1538-4357/acd337}

\bibitem[{{Bonamigo} {et~al.}(2015){Bonamigo}, {Despali}, {Limousin}, {Angulo},
  {Giocoli}, \& {Soucail}}]{bonamigo}
{Bonamigo}, M., {Despali}, G., {Limousin}, M., {et~al.} 2015, \mnras, 449,
  3171, \dodoi{10.1093/mnras/stv417}

\bibitem[{{Buck} {et~al.}(2018){Buck}, {Macci{\`o}}, {Ness}, {Obreja}, \&
  {Dutton}}]{nihao1}
{Buck}, T., {Macci{\`o}}, A., {Ness}, M., {Obreja}, A., \& {Dutton}, A. 2018,
  in Rediscovering Our Galaxy, ed. C.~{Chiappini}, I.~{Minchev},
  E.~{Starkenburg}, \& M.~{Valentini}, Vol. 334, 209--212,
  \dodoi{10.1017/S1743921317006573}

\bibitem[{{Buck} {et~al.}(2020){Buck}, {Obreja}, {Macci{\`o}}, {Minchev},
  {Dutton}, \& {Ostriker}}]{nihao2}
{Buck}, T., {Obreja}, A., {Macci{\`o}}, A.~V., {et~al.} 2020, \mnras, 491,
  3461, \dodoi{10.1093/mnras/stz3241}

\bibitem[{{Cappellari} {et~al.}(2011){Cappellari}, {Emsellem}, {Krajnovi{\'c}},
  {McDermid}, {Serra}, {Alatalo}, {Blitz}, {Bois}, {Bournaud}, {Bureau},
  {Davies}, {Davis}, {de Zeeuw}, {Khochfar}, {Kuntschner}, {Lablanche},
  {Morganti}, {Naab}, {Oosterloo}, {Sarzi}, {Scott}, {Weijmans}, \&
  {Young}}]{hubblecomb}
{Cappellari}, M., {Emsellem}, E., {Krajnovi{\'c}}, D., {et~al.} 2011, \mnras,
  416, 1680, \dodoi{10.1111/j.1365-2966.2011.18600.x}

\bibitem[{{Caputo}(1998)}]{PopII}
{Caputo}, F. 1998, \aapr, 9, 33, \dodoi{10.1007/s001590050014}

\bibitem[{Casares(2001)}]{HII_recombination}
Casares, J. 2001, X-RayBinaries and Black Hole Candidates: A Review of Optical
  Properties (Berlin, Heidelberg: Springer Berlin Heidelberg), 277--327,
  \dodoi{10.1007/3-540-44395-9_6}

\bibitem[{{Cen} \& {Ostriker}(1999)}]{cen_ostriker99}
{Cen}, R., \& {Ostriker}, J.~P. 1999, \apj, 514, 1, \dodoi{10.1086/306949}

\bibitem[{{Corbelli} {et~al.}(2014){Corbelli}, {Thilker}, {Zibetti},
  {Giovanardi}, \& {Salucci}}]{m3321cm}
{Corbelli}, E., {Thilker}, D., {Zibetti}, S., {Giovanardi}, C., \& {Salucci},
  P. 2014, \aap, 572, A23, \dodoi{10.1051/0004-6361/201424033}

\bibitem[{{Crain} {et~al.}(2015){Crain}, {Schaye}, {Bower}, {Furlong},
  {Schaller}, {Theuns}, {Dalla Vecchia}, {Frenk}, {McCarthy}, {Helly},
  {Jenkins}, {Rosas-Guevara}, {White}, \& {Trayford}}]{eagle1}
{Crain}, R.~A., {Schaye}, J., {Bower}, R.~G., {et~al.} 2015, \mnras, 450, 1937,
  \dodoi{10.1093/mnras/stv725}

\bibitem[{{Dav{\'e}} {et~al.}(2001){Dav{\'e}}, {Cen}, {Ostriker}, {Bryan},
  {Hernquist}, {Katz}, {Weinberg}, {Norman}, \& {O'Shea}}]{IGM}
{Dav{\'e}}, R., {Cen}, R., {Ostriker}, J.~P., {et~al.} 2001, \apj, 552, 473,
  \dodoi{10.1086/320548}

\bibitem[{{de Vaucouleurs}(1959)}]{DeVal}
{de Vaucouleurs}, G. 1959, Handbuch der Physik, 53, 275,
  \dodoi{10.1007/978-3-642-45932-0\_7}

\bibitem[{{El-Badry} {et~al.}(2018){El-Badry}, {Quataert}, {Wetzel}, {Hopkins},
  {Weisz}, {Chan}, {Fitts}, {Boylan-Kolchin}, {Kere{\v{s}}},
  {Faucher-Gigu{\`e}re}, \& {Garrison-Kimmel}}]{fire3}
{El-Badry}, K., {Quataert}, E., {Wetzel}, A., {et~al.} 2018, \mnras, 473, 1930,
  \dodoi{10.1093/mnras/stx2482}

\bibitem[{{Emami} {et~al.}(2020){Emami}, {Hernquist}, {Alcock}, {Genel},
  {Bose}, {Weinberger}, {Vogelsberger}, {Shen}, {Speagle}, {Marinacci},
  {Forbes}, \& {Torrey}}]{StarMorph}
{Emami}, R., {Hernquist}, L., {Alcock}, C., {et~al.} 2020, arXiv e-prints,
  arXiv:2012.12284.
\newblock \doarXiv{2012.12284}

\bibitem[{{Emami} {et~al.}(2021){Emami}, {Genel}, {Hernquist}, {Alcock},
  {Bose}, {Weinberger}, {Vogelsberger}, {Marinacci}, {Loeb}, {Torrey}, \&
  {Forbes}}]{DMHaloMorph}
{Emami}, R., {Genel}, S., {Hernquist}, L., {et~al.} 2021, \apj, 913, 36,
  \dodoi{10.3847/1538-4357/abf147}

\bibitem[{{Emami} {et~al.}(2022){Emami}, {Hernquist}, {Vogelsberger}, {Shen},
  {Speagle}, {Moreno}, {Alcock}, {Genel}, {Forbes}, {Marinacci}, \&
  {Torrey}}]{2022arXiv220207162E}
{Emami}, R., {Hernquist}, L., {Vogelsberger}, M., {et~al.} 2022, arXiv
  e-prints, arXiv:2202.07162.
\newblock \doarXiv{2202.07162}

\bibitem[{{Foster} {et~al.}(2012){Foster}, {Ji}, {Smith}, \&
  {Brickhouse}}]{Foster2012}
{Foster}, A.~R., {Ji}, L., {Smith}, R.~K., \& {Brickhouse}, N.~S. 2012, \apj,
  756, 128, \dodoi{10.1088/0004-637X/756/2/128}

\bibitem[{{Garrison-Kimmel} {et~al.}(2018){Garrison-Kimmel}, {Hopkins},
  {Wetzel}, {El-Badry}, {Sanderson}, {Bullock}, {Ma}, {van de Voort}, {Hafen},
  {Faucher-Gigu{\`e}re}, {Hayward}, {Quataert}, {Kere{\v{s}}}, \&
  {Boylan-Kolchin}}]{fire2}
{Garrison-Kimmel}, S., {Hopkins}, P.~F., {Wetzel}, A., {et~al.} 2018, \mnras,
  481, 4133, \dodoi{10.1093/mnras/sty2513}

\bibitem[{{Grand} {et~al.}(2018){Grand}, {Helly}, {Fattahi}, {Cautun}, {Cole},
  {Cooper}, {Deason}, {Frenk}, {G{\'o}mez}, {Hunt}, {Marinacci}, {Pakmor},
  {Simpson}, {Springel}, \& {Xu}}]{auriga2}
{Grand}, R. J.~J., {Helly}, J., {Fattahi}, A., {et~al.} 2018, \mnras, 481,
  1726, \dodoi{10.1093/mnras/sty2403}

\bibitem[{{Hafen} {et~al.}(2019){Hafen}, {Faucher-Gigu{\`e}re},
  {Angl{\'e}s-Alc{\'a}zar}, {Stern}, {Kere{\v{s}}}, {Hummels}, {Esmerian},
  {Garrison-Kimmel}, {El-Badry}, {Wetzel}, {Chan}, {Hopkins}, \&
  {Murray}}]{2019MNRAS.488.1248H}
{Hafen}, Z., {Faucher-Gigu{\`e}re}, C.-A., {Angl{\'e}s-Alc{\'a}zar}, D.,
  {et~al.} 2019, \mnras, 488, 1248, \dodoi{10.1093/mnras/stz1773}

\bibitem[{{Hafen} {et~al.}(2020){Hafen}, {Faucher-Gigu{\`e}re},
  {Angl{\'e}s-Alc{\'a}zar}, {Stern}, {Kere{\v{s}}}, {Esmerian}, {Wetzel},
  {El-Badry}, {Chan}, \& {Murray}}]{2020MNRAS.494.3581H}
---. 2020, \mnras, 494, 3581, \dodoi{10.1093/mnras/staa902}

\bibitem[{{Hani} {et~al.}(2019){Hani}, {Ellison}, {Sparre}, {Grand}, {Pakmor},
  {Gomez}, \& {Springel}}]{auriga3}
{Hani}, M.~H., {Ellison}, S.~L., {Sparre}, M., {et~al.} 2019, \mnras, 488, 135,
  \dodoi{10.1093/mnras/stz1708}

\bibitem[{{Harris} {et~al.}(2020){Harris}, {Jarrod Millman}, {van der Walt},
  {Gommers}, {Virtanen}, {Cournapeau}, {Wieser}, {Taylor}, {Berg}, {Smith},
  {Kern}, {Picus}, {Hoyer}, {van Kerkwijk}, {Brett}, {Haldane}, {Fern{\'a}ndez
  del R{\'\i}o}, {Wiebe}, {Peterson}, {G{\'e}rard-Marchant}, {Sheppard},
  {Reddy}, {Weckesser}, {Abbasi}, {Gohlke}, \& {Oliphant}}]{numpy}
{Harris}, C.~R., {Jarrod Millman}, K., {van der Walt}, S.~J., {et~al.} 2020,
  arXiv e-prints, arXiv:2006.10256,
  \dodoi{https://doi.org/10.1038/s41586-020-2649-2}

\bibitem[{Hartwig {et~al.}(2018)Hartwig, Volonteri, \&
  Dashyan}]{10.1093/mnras/sty229}
Hartwig, T., Volonteri, M., \& Dashyan, G. 2018, Monthly Notices of the Royal
  Astronomical Society, 476, 2288, \dodoi{10.1093/mnras/sty229}

\bibitem[{{Herrera-Camus, R.} {et~al.}(2020){Herrera-Camus, R.}, {Janssen, A.},
  {Sturm, E.}, {Lutz, D.}, {Veilleux, S.}, {Davies, R.}, {Shimizu, T.},
  {Gonz\'alez-Alfonso, E.}, {Rupke, D. S. N.}, {Tacconi, L.}, {Genzel, R.},
  {Cicone, C.}, {Maiolino, R.}, {Contursi, A.}, \& {Graci\'a-Carpio,
  J.}}]{refId0}
{Herrera-Camus, R.}, {Janssen, A.}, {Sturm, E.}, {et~al.} 2020, A\&A, 635, A47,
  \dodoi{10.1051/0004-6361/201936434}

\bibitem[{{Hubble}(1926{\natexlab{a}})}]{Hubble1}
{Hubble}, E. 1926{\natexlab{a}}, Contributions from the Mount Wilson
  Observatory / Carnegie Institution of Washington, 324, 1

\bibitem[{{Hubble}(1926{\natexlab{b}})}]{Hubble2}
{Hubble}, E.~P. 1926{\natexlab{b}}, \apj, 64, 321, \dodoi{10.1086/143018}

\bibitem[{{Hubble}(1927)}]{Hubble3}
---. 1927, The Observatory, 50, 276

\bibitem[{Hubble(1982)}]{Hubble4}
Hubble, E.~P. 1982, The realm of the nebulae (Yale University Press)

\bibitem[{{Huber} \& {Bregman}(2022)}]{HB2022}
{Huber}, M.~C., \& {Bregman}, J.~N. 2022, \aj, 163, 264,
  \dodoi{10.3847/1538-3881/ac502c}

\bibitem[{{Humphrey} {et~al.}(2011){Humphrey}, {Buote}, {Canizares}, {Fabian},
  \& {Miller}}]{xray1}
{Humphrey}, P.~J., {Buote}, D.~A., {Canizares}, C.~R., {Fabian}, A.~C., \&
  {Miller}, J.~M. 2011, \apj, 729, 53, \dodoi{10.1088/0004-637X/729/1/53}

\bibitem[{{Hunter}(2007)}]{matplotlib}
{Hunter}, J.~D. 2007, Computing in Science Engineering, 9, 90

\bibitem[{{Kauffmann} {et~al.}(2019){Kauffmann}, {Nelson}, {Borthakur},
  {Heckman}, {Hernquist}, {Marinacci}, {Pakmor}, \&
  {Pillepich}}]{CGM_morphology}
{Kauffmann}, G., {Nelson}, D., {Borthakur}, S., {et~al.} 2019, \mnras, 486,
  4686, \dodoi{10.1093/mnras/stz1029}

\bibitem[{{Kraft} {et~al.}(2022){Kraft}, {Markevitch}, {Kilbourne}, {Adams},
  {Akamatsu}, {Ayromlou}, {Bandler}, {Barbera}, {Bennett}, {Bhardwaj}, {Biffi},
  {Bodewits}, {Bogdan}, {Bonamente}, {Borgani}, {Branduardi-Raymont},
  {Bregman}, {Burchett}, {Cann}, {Carter}, {Chakraborty}, {Churazov}, {Crain},
  {Cumbee}, {Dave}, {DiPirro}, {Dolag}, {Bertrand Doriese}, {Drake}, {Dunn},
  {Eckart}, {Eckert}, {Ettori}, {Forman}, {Galeazzi}, {Gall}, {Gatuzz}, {Hell},
  {Hodges-Kluck}, {Jackman}, {Jahromi}, {Jennings}, {Jones}, {Kaaret},
  {Kavanagh}, {Kelley}, {Khabibullin}, {Kim}, {Koutroumpa}, {Kovacs}, {Kuntz},
  {Lau}, {Lee}, {Leutenegger}, {Lin}, {Lisse}, {Lo Cicero}, {Lovisari},
  {McCammon}, {McEntee}, {Mernier}, {Miller}, {Nagai}, {Negro}, {Nelson},
  {Ness}, {Nulsen}, {Ogorzalek}, {Oppenheimer}, {Oskinova}, {Patnaude},
  {Pfeifle}, {Pillepich}, {Plucinsky}, {Pooley}, {Porter}, {Randall}, {Rasia},
  {Raymond}, {Ruszkowski}, {Sakai}, {Sarkar}, {Sasaki}, {Sato},
  {Schellenberger}, {Schaye}, {Simionescu}, {Smith}, {Steiner}, {Stern}, {Su},
  {Sun}, {Tremblay}, {Truong}, {Tutt}, {Ursino}, {Veilleux}, {Vikhlinin},
  {Vladutescu-Zopp}, {Vogelsberger}, {Walker}, {Weaver}, {Weigt}, {Werk},
  {Werner}, {Wolk}, {Zhang}, {Zhang}, {Zhuravleva}, \& {ZuHone}}]{LEM}
{Kraft}, R., {Markevitch}, M., {Kilbourne}, C., {et~al.} 2022, arXiv e-prints,
  arXiv:2211.09827, \dodoi{10.48550/arXiv.2211.09827}

\bibitem[{{Li} {et~al.}(2021){Li}, {Rahman}, {Murray}, {Hafen},
  {Faucher-Gigu{\`e}re}, {Stern}, {Hummels}, {Hopkins}, {El-Badry}, \&
  {Kere{\v{s}}}}]{2021MNRAS.500.1038L}
{Li}, F., {Rahman}, M., {Murray}, N., {et~al.} 2021, \mnras, 500, 1038,
  \dodoi{10.1093/mnras/staa3322}

\bibitem[{{Li} {et~al.}(2017){Li}, {Bregman}, {Wang}, {Crain}, {Anderson}, \&
  {Zhang}}]{Tvir}
{Li}, J.-T., {Bregman}, J.~N., {Wang}, Q.~D., {et~al.} 2017, \apjs, 233, 20,
  \dodoi{10.3847/1538-4365/aa96fc}

\bibitem[{{Liang} {et~al.}(2018){Liang}, {Kravtsov}, \& {Agertz}}]{CGM}
{Liang}, C.~J., {Kravtsov}, A.~V., \& {Agertz}, O. 2018, \mnras, 479, 1822,
  \dodoi{10.1093/mnras/sty1668}

\bibitem[{{Mantz} {et~al.}(2017){Mantz}, {Allen}, {Morris}, {Simionescu},
  {Urban}, {Werner}, \& {Zhuravleva}}]{Mantz2017}
{Mantz}, A.~B., {Allen}, S.~W., {Morris}, R.~G., {et~al.} 2017, \mnras, 472,
  2877, \dodoi{10.1093/mnras/stx2200}

\bibitem[{{Marinacci} {et~al.}(2018){Marinacci}, {Vogelsberger}, {Pakmor},
  {Torrey}, {Springel}, {Hernquist}, {Nelson}, {Weinberger}, {Pillepich},
  {Naiman}, \& {Genel}}]{tng3}
{Marinacci}, F., {Vogelsberger}, M., {Pakmor}, R., {et~al.} 2018, \mnras, 480,
  5113, \dodoi{10.1093/mnras/sty2206}

\bibitem[{{Mathur} {et~al.}(2023){Mathur}, {Das}, {Gupta}, \&
  {Krongold}}]{warmhot}
{Mathur}, S., {Das}, S., {Gupta}, A., \& {Krongold}, Y. 2023, \mnras, 525, L11,
  \dodoi{10.1093/mnrasl/slad085}

\bibitem[{{McDonald} {et~al.}(2016){McDonald}, {Bulbul}, {de Haan}, {Miller},
  {Benson}, {Bleem}, {Brodwin}, {Carlstrom}, {Chiu}, {Forman},
  {Hlavacek-Larrondo}, {Garmire}, {Gupta}, {Mohr}, {Reichardt}, {Saro},
  {Stalder}, {Stark}, \& {Vieira}}]{McDonald2016}
{McDonald}, M., {Bulbul}, E., {de Haan}, T., {et~al.} 2016, \apj, 826, 124,
  \dodoi{10.3847/0004-637X/826/2/124}

\bibitem[{{Merritt} {et~al.}(2020){Merritt}, {Pillepich}, {van Dokkum},
  {Nelson}, {Hernquist}, {Marinacci}, \& {Vogelsberger}}]{tng7}
{Merritt}, A., {Pillepich}, A., {van Dokkum}, P., {et~al.} 2020, \mnras, 495,
  4570, \dodoi{10.1093/mnras/staa1164}

\bibitem[{{Miller} \& {Bregman}(2013)}]{structure_MW_halo}
{Miller}, M.~J., \& {Bregman}, J.~N. 2013, \apj, 770, 118,
  \dodoi{10.1088/0004-637X/770/2/118}

\bibitem[{{Miller} \& {Bregman}(2015)}]{MW_halo_emission}
---. 2015, \apj, 800, 14, \dodoi{10.1088/0004-637X/800/1/14}

\bibitem[{{Monachesi} {et~al.}(2016){Monachesi}, {G{\'o}mez}, {Grand},
  {Kauffmann}, {Marinacci}, {Pakmor}, {Springel}, \& {Frenk}}]{auriga1}
{Monachesi}, A., {G{\'o}mez}, F.~A., {Grand}, R. J.~J., {et~al.} 2016, \mnras,
  459, L46, \dodoi{10.1093/mnrasl/slw052}

\bibitem[{{Morrison} \& {McCammon}(1983)}]{Morrison1983}
{Morrison}, R., \& {McCammon}, D. 1983, \apj, 270, 119, \dodoi{10.1086/161102}

\bibitem[{{Naiman} {et~al.}(2018){Naiman}, {Pillepich}, {Springel},
  {Ramirez-Ruiz}, {Torrey}, {Vogelsberger}, {Pakmor}, {Nelson}, {Marinacci},
  {Hernquist}, {Weinberger}, \& {Genel}}]{tng1}
{Naiman}, J.~P., {Pillepich}, A., {Springel}, V., {et~al.} 2018, \mnras, 477,
  1206, \dodoi{10.1093/mnras/sty618}

\bibitem[{{Nakashima} {et~al.}(2018){Nakashima}, {Inoue}, {Yamasaki}, {Sofue},
  {Kataoka}, \& {Sakai}}]{MW_gas_spacial_distrib}
{Nakashima}, S., {Inoue}, Y., {Yamasaki}, N., {et~al.} 2018, \apj, 862, 34,
  \dodoi{10.3847/1538-4357/aacceb}

\bibitem[{{Nelson} {et~al.}(2018){Nelson}, {Kauffmann}, {Pillepich}, {Genel},
  {Springel}, {Pakmor}, {Hernquist}, {Weinberger}, {Torrey}, {Vogelsberger}, \&
  {Marinacci}}]{tng2}
{Nelson}, D., {Kauffmann}, G., {Pillepich}, A., {et~al.} 2018, \mnras, 477,
  450, \dodoi{10.1093/mnras/sty656}

\bibitem[{{Nelson} {et~al.}(2019{\natexlab{a}}){Nelson}, {Springel},
  {Pillepich}, {Rodriguez-Gomez}, {Torrey}, {Genel}, {Vogelsberger}, {Pakmor},
  {Marinacci}, {Weinberger}, {Kelley}, {Lovell}, {Diemer}, \&
  {Hernquist}}]{TNGDataRelease}
{Nelson}, D., {Springel}, V., {Pillepich}, A., {et~al.} 2019{\natexlab{a}},
  Computational Astrophysics and Cosmology, 6, 2,
  \dodoi{10.1186/s40668-019-0028-x}

\bibitem[{{Nelson} {et~al.}(2019{\natexlab{b}}){Nelson}, {Pillepich},
  {Springel}, {Pakmor}, {Weinberger}, {Genel}, {Torrey}, {Vogelsberger},
  {Marinacci}, \& {Hernquist}}]{TNGFirstResultsFB}
{Nelson}, D., {Pillepich}, A., {Springel}, V., {et~al.} 2019{\natexlab{b}},
  \mnras, 490, 3234, \dodoi{10.1093/mnras/stz2306}

\bibitem[{{Oliphant}(2007)}]{2007CSE.....9c..10O}
{Oliphant}, T.~E. 2007, Computing in Science and Engineering, 9, 10,
  \dodoi{10.1109/MCSE.2007.58}

\bibitem[{{Orr} {et~al.}(2020){Orr}, {Hayward}, {Medling}, {Gurvich},
  {Hopkins}, {Murray}, {Pineda}, {Faucher-Gigu{\`e}re}, {Kere{\v{s}}},
  {Wetzel}, \& {Su}}]{fire4}
{Orr}, M.~E., {Hayward}, C.~C., {Medling}, A.~M., {et~al.} 2020, \mnras, 496,
  1620, \dodoi{10.1093/mnras/staa1619}

\bibitem[{{Pillepich} {et~al.}(2018{\natexlab{a}}){Pillepich}, {Nelson},
  {Hernquist}, {Springel}, {Pakmor}, {Torrey}, {Weinberger}, {Genel}, {Naiman},
  {Marinacci}, \& {Vogelsberger}}]{unresolved}
{Pillepich}, A., {Nelson}, D., {Hernquist}, L., {et~al.} 2018{\natexlab{a}},
  \mnras, 475, 648, \dodoi{10.1093/mnras/stx3112}

\bibitem[{{Pillepich} {et~al.}(2018{\natexlab{b}}){Pillepich}, {Nelson},
  {Hernquist}, {Springel}, {Pakmor}, {Torrey}, {Weinberger}, {Genel}, {Naiman},
  {Marinacci}, \& {Vogelsberger}}]{tng4}
---. 2018{\natexlab{b}}, \mnras, 475, 648, \dodoi{10.1093/mnras/stx3112}

\bibitem[{{Pillepich} {et~al.}(2019){Pillepich}, {Nelson}, {Springel},
  {Pakmor}, {Torrey}, {Weinberger}, {Vogelsberger}, {Marinacci}, {Genel}, {van
  der Wel}, \& {Hernquist}}]{TNGFirstResultsEV}
{Pillepich}, A., {Nelson}, D., {Springel}, V., {et~al.} 2019, \mnras, 490,
  3196, \dodoi{10.1093/mnras/stz2338}

\bibitem[{{Planck Collaboration} {et~al.}(2016){Planck Collaboration}, {Ade},
  {Aghanim}, {Arnaud}, {Ashdown}, {Aumont}, {Baccigalupi}, {Banday},
  {Barreiro}, {Bartlett}, {Bartolo}, {Battaner}, {Battye}, {Benabed},
  {Beno{\^\i}t}, {Benoit-L{\'e}vy}, {Bernard}, {Bersanelli}, {Bielewicz},
  {Bock}, {Bonaldi}, {Bonavera}, {Bond}, {Borrill}, {Bouchet}, {Boulanger},
  {Bucher}, {Burigana}, {Butler}, {Calabrese}, {Cardoso}, {Catalano},
  {Challinor}, {Chamballu}, {Chary}, {Chiang}, {Chluba}, {Christensen},
  {Church}, {Clements}, {Colombi}, {Colombo}, {Combet}, {Coulais}, {Crill},
  {Curto}, {Cuttaia}, {Danese}, {Davies}, {Davis}, {de Bernardis}, {de Rosa},
  {de Zotti}, {Delabrouille}, {D{\'e}sert}, {Di Valentino}, {Dickinson},
  {Diego}, {Dolag}, {Dole}, {Donzelli}, {Dor{\'e}}, {Douspis}, {Ducout},
  {Dunkley}, {Dupac}, {Efstathiou}, {Elsner}, {En{\ss}lin}, {Eriksen},
  {Farhang}, {Fergusson}, {Finelli}, {Forni}, {Frailis}, {Fraisse},
  {Franceschi}, {Frejsel}, {Galeotta}, {Galli}, {Ganga}, {Gauthier}, {Gerbino},
  {Ghosh}, {Giard}, {Giraud-H{\'e}raud}, {Giusarma}, {Gjerl{\o}w},
  {Gonz{\'a}lez-Nuevo}, {G{\'o}rski}, {Gratton}, {Gregorio}, {Gruppuso},
  {Gudmundsson}, {Hamann}, {Hansen}, {Hanson}, {Harrison}, {Helou},
  {Henrot-Versill{\'e}}, {Hern{\'a}ndez-Monteagudo}, {Herranz}, {Hildebrandt},
  {Hivon}, {Hobson}, {Holmes}, {Hornstrup}, {Hovest}, {Huang}, {Huffenberger},
  {Hurier}, {Jaffe}, {Jaffe}, {Jones}, {Juvela}, {Keih{\"a}nen}, {Keskitalo},
  {Kisner}, {Kneissl}, {Knoche}, {Knox}, {Kunz}, {Kurki-Suonio}, {Lagache},
  {L{\"a}hteenm{\"a}ki}, {Lamarre}, {Lasenby}, {Lattanzi}, {Lawrence}, {Leahy},
  {Leonardi}, {Lesgourgues}, {Levrier}, {Lewis}, {Liguori}, {Lilje},
  {Linden-V{\o}rnle}, {L{\'o}pez-Caniego}, {Lubin}, {Mac{\'\i}as-P{\'e}rez},
  {Maggio}, {Maino}, {Mandolesi}, {Mangilli}, {Marchini}, {Maris}, {Martin},
  {Martinelli}, {Mart{\'\i}nez-Gonz{\'a}lez}, {Masi}, {Matarrese}, {McGehee},
  {Meinhold}, {Melchiorri}, {Melin}, {Mendes}, {Mennella}, {Migliaccio},
  {Millea}, {Mitra}, {Miville-Desch{\^e}nes}, {Moneti}, {Montier}, {Morgante},
  {Mortlock}, {Moss}, {Munshi}, {Murphy}, {Naselsky}, {Nati}, {Natoli},
  {Netterfield}, {N{\o}rgaard-Nielsen}, {Noviello}, {Novikov}, {Novikov},
  {Oxborrow}, {Paci}, {Pagano}, {Pajot}, {Paladini}, {Paoletti}, {Partridge},
  {Pasian}, {Patanchon}, {Pearson}, {Perdereau}, {Perotto}, {Perrotta},
  {Pettorino}, {Piacentini}, {Piat}, {Pierpaoli}, {Pietrobon}, {Plaszczynski},
  {Pointecouteau}, {Polenta}, {Popa}, {Pratt}, {Pr{\'e}zeau}, {Prunet},
  {Puget}, {Rachen}, {Reach}, {Rebolo}, {Reinecke}, {Remazeilles}, {Renault},
  {Renzi}, {Ristorcelli}, {Rocha}, {Rosset}, {Rossetti}, {Roudier},
  {Rouill{\'e} d'Orfeuil}, {Rowan-Robinson}, {Rubi{\~n}o-Mart{\'\i}n},
  {Rusholme}, {Said}, {Salvatelli}, {Salvati}, {Sandri}, {Santos},
  {Savelainen}, {Savini}, {Scott}, {Seiffert}, {Serra}, {Shellard}, {Spencer},
  {Spinelli}, {Stolyarov}, {Stompor}, {Sudiwala}, {Sunyaev}, {Sutton},
  {Suur-Uski}, {Sygnet}, {Tauber}, {Terenzi}, {Toffolatti}, {Tomasi},
  {Tristram}, {Trombetti}, {Tucci}, {Tuovinen}, {T{\"u}rler}, {Umana},
  {Valenziano}, {Valiviita}, {Van Tent}, {Vielva}, {Villa}, {Wade}, {Wandelt},
  {Wehus}, {White}, {White}, {Wilkinson}, {Yvon}, {Zacchei}, \&
  {Zonca}}]{TNG_parameters}
{Planck Collaboration}, {Ade}, P.~A.~R., {Aghanim}, N., {et~al.} 2016, \aap,
  594, A13, \dodoi{10.1051/0004-6361/201525830}

\bibitem[{{Pulsoni} {et~al.}(2020){Pulsoni}, {Gerhard}, {Arnaboldi},
  {Pillepich}, {Nelson}, {Hernquist}, \& {Springel}}]{pulsoni}
{Pulsoni}, C., {Gerhard}, O., {Arnaboldi}, M., {et~al.} 2020, \aap, 641, A60,
  \dodoi{10.1051/0004-6361/202038253}

\bibitem[{{Putman} {et~al.}(2012){Putman}, {Peek}, \& {Joung}}]{Emission_map1}
{Putman}, M.~E., {Peek}, J.~E.~G., \& {Joung}, M.~R. 2012, \araa, 50, 491,
  \dodoi{10.1146/annurev-astro-081811-125612}

\bibitem[{{Sanderson} {et~al.}(2020){Sanderson}, {Wetzel}, {Loebman}, {Sharma},
  {Hopkins}, {Garrison-Kimmel}, {Faucher-Gigu{\`e}re}, {Kere{\v{s}}}, \&
  {Quataert}}]{fire5}
{Sanderson}, R.~E., {Wetzel}, A., {Loebman}, S., {et~al.} 2020, \apjs, 246, 6,
  \dodoi{10.3847/1538-4365/ab5b9d}

\bibitem[{{Santistevan} {et~al.}(2020){Santistevan}, {Wetzel}, {El-Badry},
  {Bland-Hawthorn}, {Boylan-Kolchin}, {Bailin}, {Faucher-Gigu{\`e}re}, \&
  {Benincasa}}]{fire1}
{Santistevan}, I.~B., {Wetzel}, A., {El-Badry}, K., {et~al.} 2020, \mnras, 497,
  747, \dodoi{10.1093/mnras/staa1923}

\bibitem[{{Savitzky} \& {Golay}(1964)}]{savgol}
{Savitzky}, A., \& {Golay}, M.~J.~E. 1964, Analytical Chemistry, 36, 1627,
  \dodoi{10.1021/ac60214a047}

\bibitem[{{Schaye} {et~al.}(2015){Schaye}, {Crain}, {Bower}, {Furlong},
  {Schaller}, {Theuns}, {Dalla Vecchia}, {Frenk}, {McCarthy}, {Helly},
  {Jenkins}, {Rosas-Guevara}, {White}, {Baes}, {Booth}, {Camps}, {Navarro},
  {Qu}, {Rahmati}, {Sawala}, {Thomas}, \& {Trayford}}]{eagle2}
{Schaye}, J., {Crain}, R.~A., {Bower}, R.~G., {et~al.} 2015, \mnras, 446, 521,
  \dodoi{10.1093/mnras/stu2058}

\bibitem[{{Smith} {et~al.}(2001){Smith}, {Brickhouse}, {Liedahl}, \&
  {Raymond}}]{Smith2001}
{Smith}, R.~K., {Brickhouse}, N.~S., {Liedahl}, D.~A., \& {Raymond}, J.~C.
  2001, \apjl, 556, L91, \dodoi{10.1086/322992}

\bibitem[{{Smith} {et~al.}(2019){Smith}, {Abraham}, {Baird}, {Bautz},
  {Bookbinder}, {Bregman}, {Brenneman}, {Brickhouse}, {Burrows}, {Burwitz},
  {Bushman}, {Canizares}, {Chakrabarty}, {Cheimets}, {Costantini}, {Dawson},
  {DeRoo}, {Falcone}, {Foster}, {Gallo}, {Grant}, {G{\"u}nther}, {Heilmann},
  {Hine}, {Huenemoerder}, {Jara}, {Kaastra}, {Kreykenbohm}, {Madsen},
  {McDonald}, {McEachen}, {McEntaffer}, {Marshall}, {Miller}, {Miller},
  {Morse}, {Mushotzky}, {Nandra}, {Nowak}, {Paerels}, {Petre}, {Poppenhaeger},
  {Ptak}, {Reid}, {Ronzano}, {Sanders}, {Schattenburg}, {Schonfeld}, {Schulz},
  {Smale}, {Temi}, {Valencic}, {Walker}, {Willingale}, {Wilms}, \&
  {Wolk}}]{smith2019}
{Smith}, R.~K., {Abraham}, M., {Baird}, G., {et~al.} 2019, in Society of
  Photo-Optical Instrumentation Engineers (SPIE) Conference Series, Vol. 11118,
  UV, X-Ray, and Gamma-Ray Space Instrumentation for Astronomy XXI, ed. O.~H.
  {Siegmund}, 111180W, \dodoi{10.1117/12.2529499}

\bibitem[{{Sparke} {et~al.}(2009){Sparke}, {van Moorsel}, {Schwarz}, \&
  {Vogelaar}}]{21cmwarpeddisk}
{Sparke}, L.~S., {van Moorsel}, G., {Schwarz}, U.~J., \& {Vogelaar}, M. 2009,
  \aj, 137, 3976, \dodoi{10.1088/0004-6256/137/4/3976}

\bibitem[{Springel(2010)}]{AREPO}
Springel, V. 2010, Monthly Notices of the Royal Astronomical Society, 401, 791,
  \dodoi{10.1111/j.1365-2966.2009.15715.x}

\bibitem[{{Springel} {et~al.}(2018){Springel}, {Pakmor}, {Pillepich},
  {Weinberger}, {Nelson}, {Hernquist}, {Vogelsberger}, {Genel}, {Torrey},
  {Marinacci}, \& {Naiman}}]{tng5}
{Springel}, V., {Pakmor}, R., {Pillepich}, A., {et~al.} 2018, \mnras, 475, 676,
  \dodoi{10.1093/mnras/stx3304}

\bibitem[{{Tchernyshyov} {et~al.}(2022){Tchernyshyov}, {Werk}, {Wilde},
  {Prochaska}, {Tripp}, {Burchett}, {Bordoloi}, {Howk}, {Lehner}, {O'Meara},
  {Tejos}, \& {Tumlinson}}]{2022ApJ...927..147T}
{Tchernyshyov}, K., {Werk}, J.~K., {Wilde}, M.~C., {et~al.} 2022, \apj, 927,
  147, \dodoi{10.3847/1538-4357/ac450c}

\bibitem[{{Trayford} {et~al.}(2019){Trayford}, {Frenk}, {Theuns}, {Schaye}, \&
  {Correa}}]{eagle4}
{Trayford}, J.~W., {Frenk}, C.~S., {Theuns}, T., {Schaye}, J., \& {Correa}, C.
  2019, \mnras, 483, 744, \dodoi{10.1093/mnras/sty2860}

\bibitem[{{Trayford} \& {Schaye}(2019)}]{eagle3}
{Trayford}, J.~W., \& {Schaye}, J. 2019, \mnras, 485, 5715,
  \dodoi{10.1093/mnras/stz757}

\bibitem[{{Tumlinson} {et~al.}(2017){Tumlinson}, {Peeples}, \&
  {Werk}}]{CGM_review}
{Tumlinson}, J., {Peeples}, M.~S., \& {Werk}, J.~K. 2017, \araa, 55, 389,
  \dodoi{10.1146/annurev-astro-091916-055240}

\bibitem[{van Rossum(1995)}]{python}
van Rossum, G. 1995, Python tutorial, Tech. Rep. CS-R9526, Centrum voor
  Wiskunde en Informatica (CWI), Amsterdam

\bibitem[{{Vogelsberger} {et~al.}(2020){Vogelsberger}, {Nelson}, {Pillepich},
  {Shen}, {Marinacci}, {Springel}, {Pakmor}, {Tacchella}, {Weinberger},
  {Torrey}, \& {Hernquist}}]{tng6}
{Vogelsberger}, M., {Nelson}, D., {Pillepich}, A., {et~al.} 2020, \mnras, 492,
  5167, \dodoi{10.1093/mnras/staa137}

\bibitem[{{Weinberger} {et~al.}(2017){Weinberger}, {Springel}, {Hernquist},
  {Pillepich}, {Marinacci}, {Pakmor}, {Nelson}, {Genel}, {Vogelsberger},
  {Naiman}, \& {Torrey}}]{Weinberger2017}
{Weinberger}, R., {Springel}, V., {Hernquist}, L., {et~al.} 2017, \mnras, 465,
  3291, \dodoi{10.1093/mnras/stw2944}

\bibitem[{{Wilde} {et~al.}(2021){Wilde}, {Werk}, {Burchett}, {Prochaska},
  {Tchernyshyov}, {Tripp}, {Tejos}, {Lehner}, {Bordoloi}, {O'Meara}, \&
  {Tumlinson}}]{2021ApJ...912....9W}
{Wilde}, M.~C., {Werk}, J.~K., {Burchett}, J.~N., {et~al.} 2021, \apj, 912, 9,
  \dodoi{10.3847/1538-4357/abea14}

\bibitem[{{Zemp} {et~al.}(2011){Zemp}, {Gnedin}, {Gnedin}, \&
  {Kravtsov}}]{zemp}
{Zemp}, M., {Gnedin}, O.~Y., {Gnedin}, N.~Y., \& {Kravtsov}, A.~V. 2011, \apjs,
  197, 30, \dodoi{10.1088/0067-0049/197/2/30}

\bibitem[{{Zhezher} {et~al.}(2016){Zhezher}, {Nugaev}, \&
  {Rubtsov}}]{MW_density_pulsars}
{Zhezher}, Y.~V., {Nugaev}, E.~Y., \& {Rubtsov}, G.~I. 2016, Astronomy Letters,
  42, 173, \dodoi{10.1134/S1063773716030063}

\end{thebibliography}
\bibliographystyle{aasjournal}

\FloatBarrier

\appendix 

\section{Concerning Arbitrary Eigenvector Orientation} \label{appendix:eigenvectors}

    When computing the radial profiles of our distributions (see Section~\ref{sec: shape_class} and Figure~\ref{fig: eigen_angle}), it was important to account for the fact that an eigenvector's orientation, in general, is not uniquely defined and can point in any of two possible directions. When computing these profiles, we begin by arbitrarily fixing the orientation of the eigenvectors corresponding to the computed eigenvalues at the innermost radius. To select the orientation of the eigenvectors for the second radial point, we compute the rotation angles between the first set of eigenvectors and the second set of eigenvectors. Since there are two possible sets of orientations for the second set of eigenvectors, we select the orientation corresponding to that which minimizes the difference in the angles between the two sets of eigenvectors. We then fix the orientation of the eigenvectors at the second radial point. This process is then repeated for the remaining $i$ sets of eigenvectors at each radius where the orientation of the $i$th eigenvector set is fixed by minimizing the rotation angles between the eigenvector set $i$ and eigenvector set $i-1$. By fixing the orientation of these eigenvectors from the minimum rotation angles, we can obtain global rotation angles of up to $180^\circ$ with respect to the fixed eigenvectors at the innermost radius.

    A major limitation of this method arises in the case of stretched distributions, where the eigenvectors display a $90^\circ$ rotation. Following a $90^\circ$, checking for the minimum angle of rotation will return $90^\circ$ for both orientations, resulting in an arbitrary orientation for this set of eigenvectors. However, LSIM is sensitive to the substructures within the gas \citep{StarMorph}. As a result, the $90^\circ$ rotations in the case of stretched distributions are not instantaneous and contain several eigenvalues throughout the rotation. Therefore, obtaining the orientation of these eigenvectors is possible with the method described above. However, in the event that an incorrect orientation is chosen (choosing $-90^\circ$ for example), the Savitzky–Golay filter (see Section~\ref{sec: shape_class}) will smooth out the resultant sharp spike in the radial profile.
    \clearpage

\section{Alternate Visualizations} \label{appendix:viz}

\begin{figure*}[ht]
\centering
\includegraphics[width = 0.8\paperwidth]{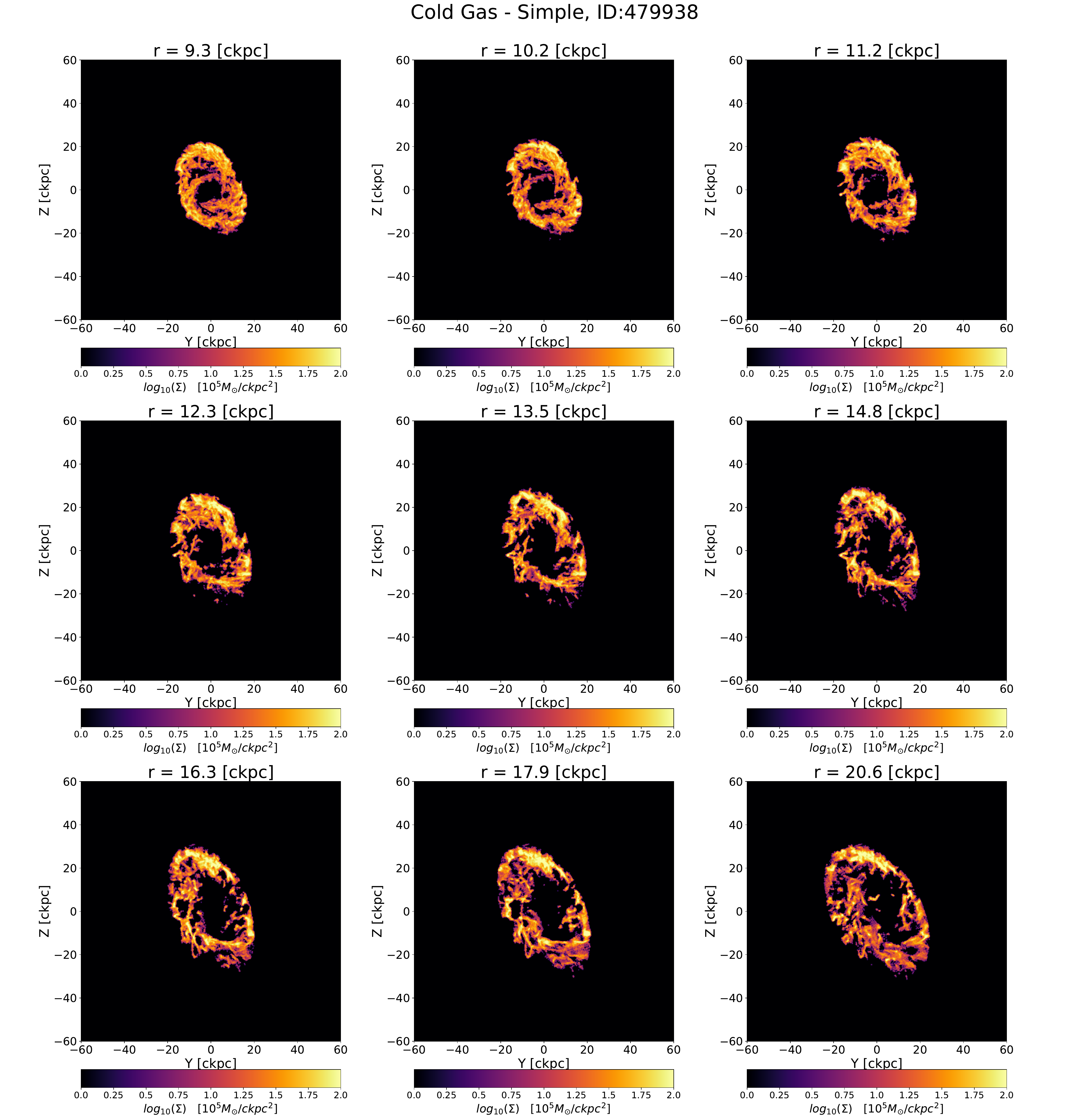}
\caption{Particle density ($\Sigma$) maps of the YZ projection for 3D shells implemented in the LSIM algorithm (see Section~\ref{sec: LSIM}) of increasing radii for a cold gas simple distribution with ID 479938 showing very little variation in axes lengths or net rotation over $11.3 \, \rm ckpc$.
\label{fig: simple_projection}}
\end{figure*}

\begin{figure*}[ht]
\centering
\includegraphics[width = 0.78\paperwidth]{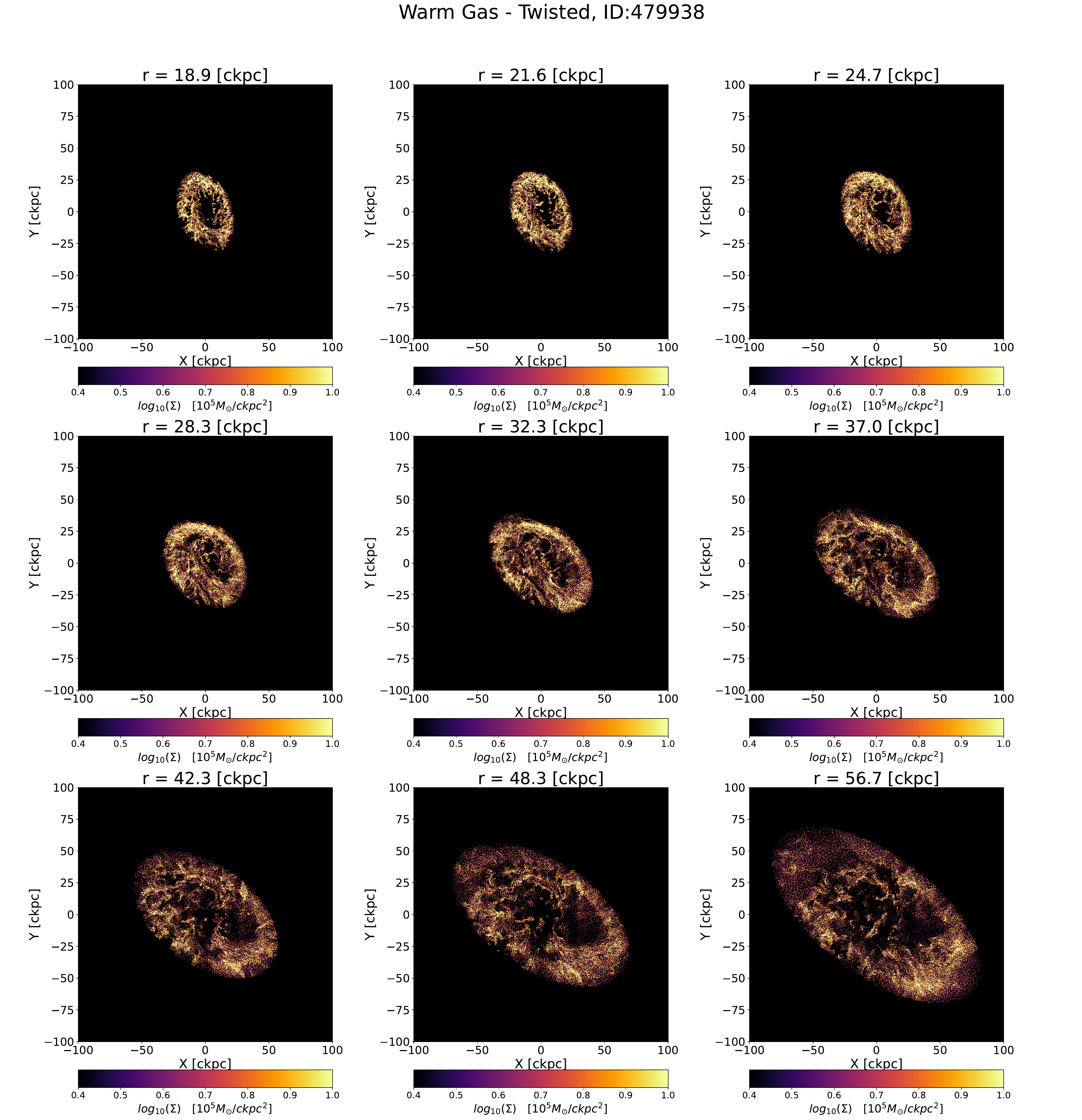}
\caption{Particle density ($\Sigma$) maps of the YZ projection for 3D shells implemented in the LSIM calculations (see Section~\ref{sec: LSIM}) of increasing radii for a warm gas twisted distribution with ID 479938 showing an approximately $20^\circ$ counterclockwise net rotation over $39.5 \, \rm ckpc$.
\label{fig: twisted_projection}}
\end{figure*}

\begin{figure*}[ht]
\centering
\includegraphics[width = 0.8\paperwidth]{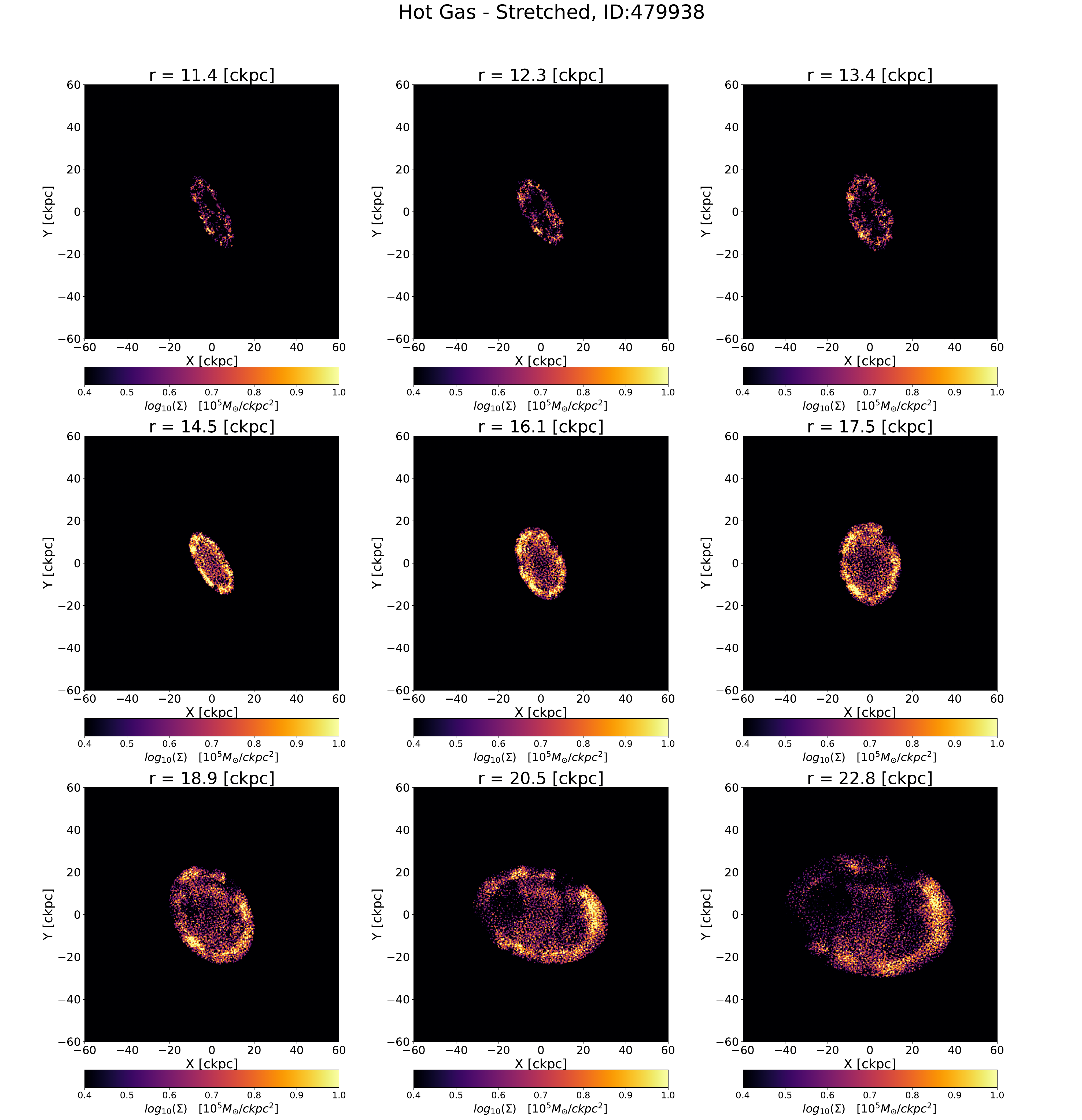}
\caption{Particle density ($\Sigma$) maps of the XY projection for 3D shells implemented in the LSIM calculations (see Section~\ref{sec: LSIM}) of increasing radii for hot gas stretched distribution with ID 479938 showing significant stretching in the minimum axis of the projected ellipse. This distribution also displays some level of counterclockwise net rotation.
\label{fig: stretched_projection}}
\end{figure*}

\begin{figure*}[ht]
\centering
\includegraphics[width = 0.8\paperwidth]{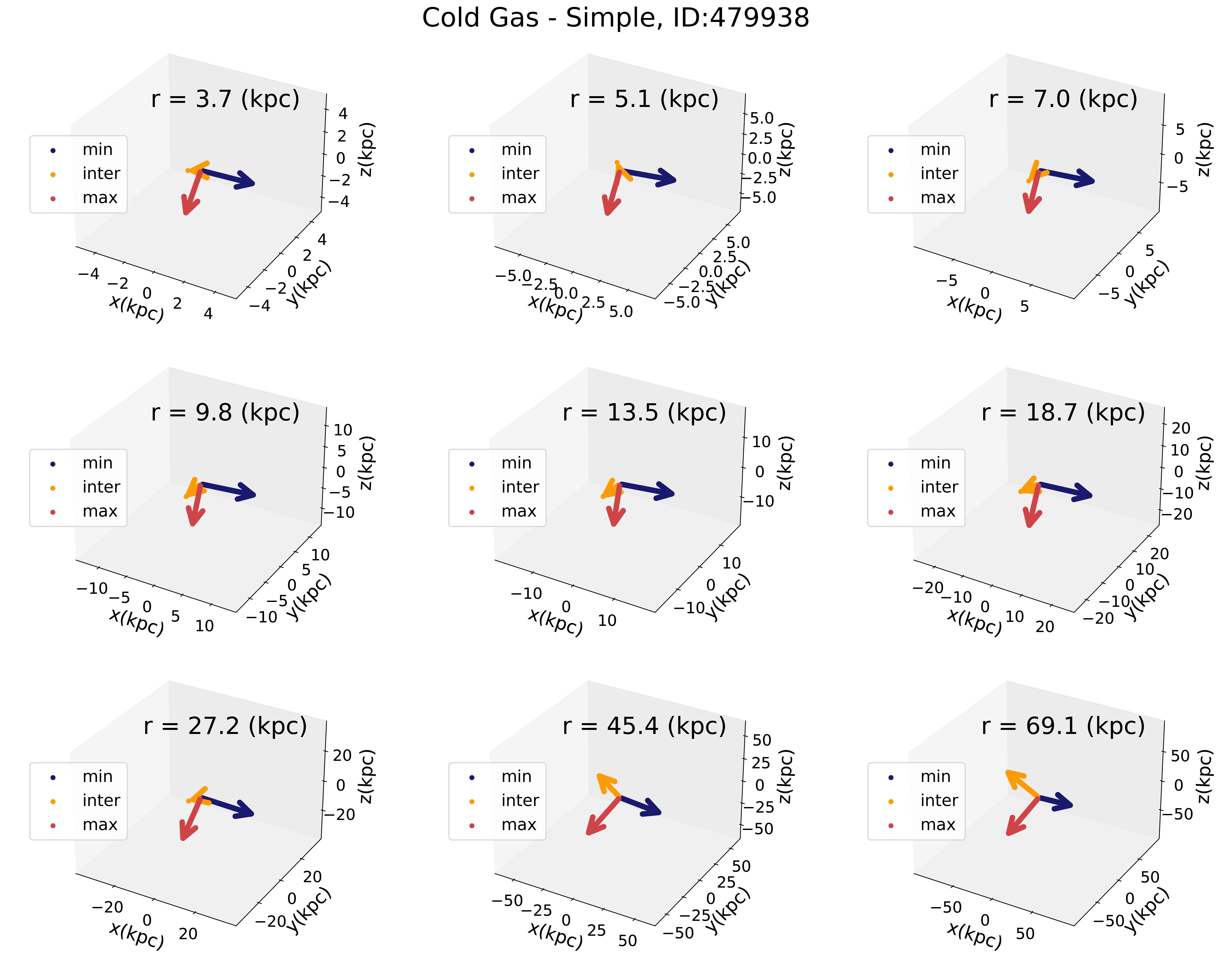}
\caption{A 3D Vector projection for the minimum, intermediate, and maximum eigenvectors of the reduced inertia tensor for a simple cold gas distribution with the ID 479938. 
\label{fig: simple_3d}}
\end{figure*}

\begin{figure*}[ht]
\centering
\includegraphics[width = 0.8\paperwidth]{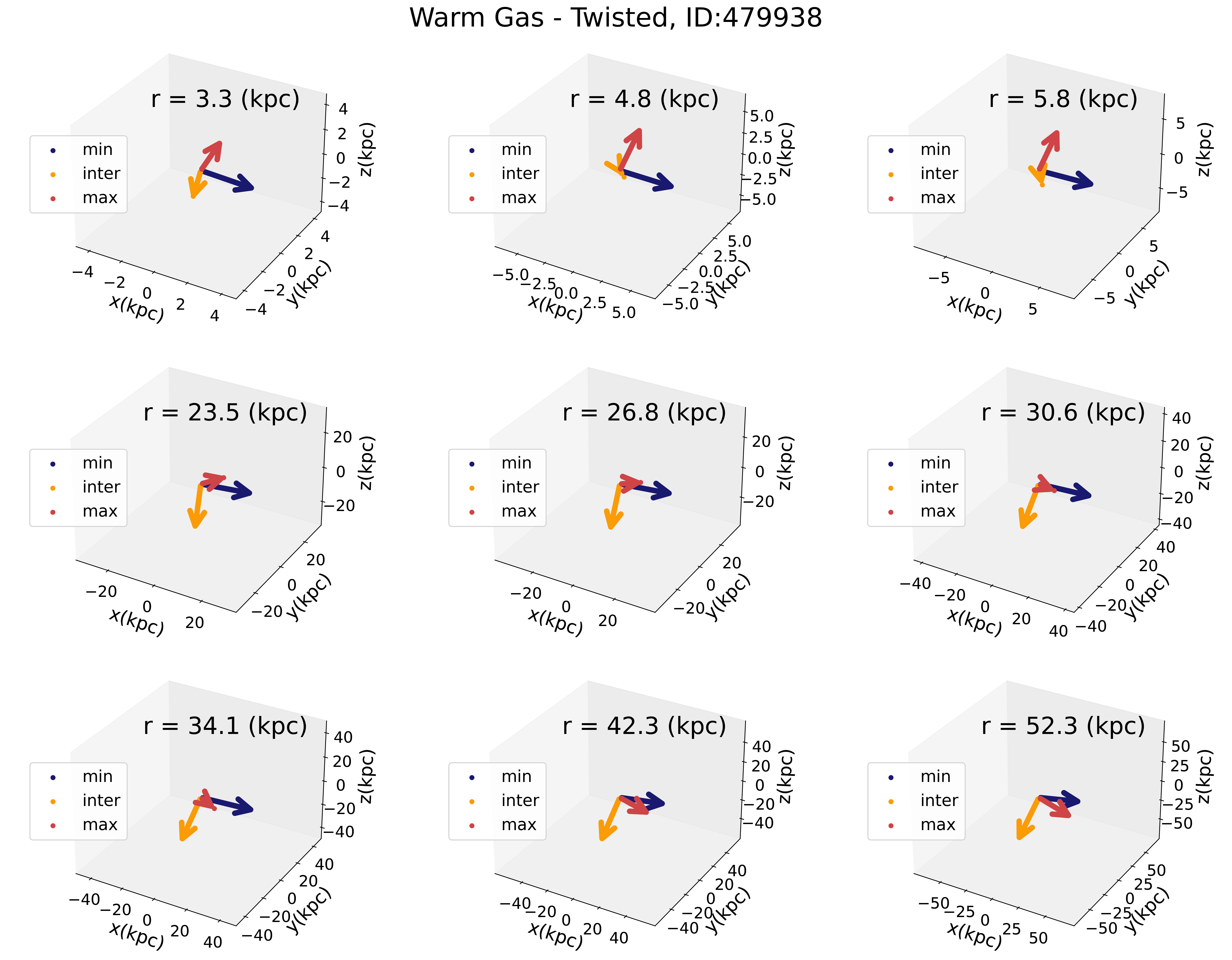}
\caption{A 3D Vector projection for the minimum, intermediate, and maximum eigenvectors of the reduced inertia tensor of the twisted warm gas distribution with the ID 479938. 
\label{fig: twisted_3d}}
\end{figure*}

\begin{figure*}[ht]
\centering
\includegraphics[width = 0.8\paperwidth]{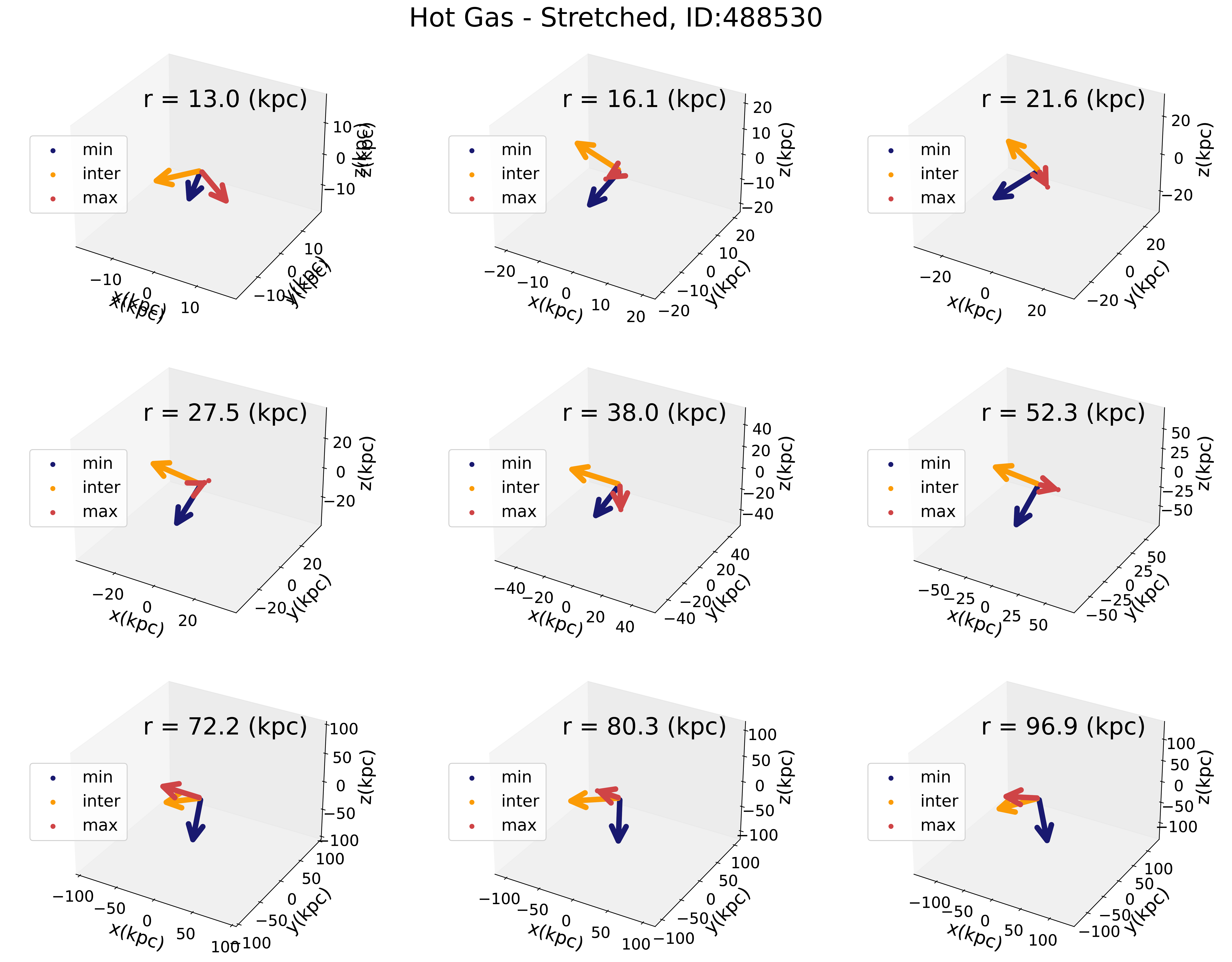}
\caption{A 3D Vector projection for the minimum, intermediate, and maximum eigenvectors of the reduced inertia tensor for the stretched hot gas distribution of the Halo with ID 488530. 
\label{fig: stretched_3d}}
\end{figure*}


\end{document}